\documentclass[authoryear, two column, 5p]{elsarticle}

\usepackage{graphicx}

\usepackage{txfonts}
\usepackage{xspace}
\usepackage{color}
\usepackage{lineno}

\usepackage{natbib}
\usepackage{geometry}
\usepackage{fleqn}

\def\mic{$\mu$m\xspace}

\def\textNew{black}
\def\refrevisions1{black}

\newcommand*\aap{A\&A}

\newcommand*\ao{Appl.~Opt.}

\newcommand*\apjl{ApJ}

\newcommand*\apjs{ApJS}

\begin{document} 

\title{Laboratory mid-IR spectra of equilibrated and igneous meteorites \\
\large{Searching for observables of planetesimal debris}}

\author[astrob,an,estec]{B.L. de Vries}
\ead{bldevries.science@gmail.com}
\author[nrm]{H. Skogby}
\author[sron,uva]{L.B.F.M. Waters}
\author[sron,uva]{M. Min}

\address[astrob]{Stockholm University Astrobiology Centre, SE-106 91 Stockholm, Sweden}
\address[an]{AlbaNova University Centre, Stockholm University, Department of Astronomy, SE-106 91 Stockholm, Sweden}
\address[estec]{Scientific Support Office, Directorate of Science, European Space Research and Technology Centre (ESA/ESTEC), Keplerlaan 1, 2201 AZ Noordwijk, The Netherlands}
\address[nrm]{Department of Geosciences, Swedish Museum of Natural History, Box 50007, SE-104 05 Stockholm, Sweden}
\address[sron]{SRON Netherlands Institute for Space Research, Sorbonnelaan 2, 3584 CA Utrecht, The Netherlands}
\address[uva]{Astronomical Institute Anton Pannekoek, University of Amsterdam, Science Park 904, 1098 XH, Amsterdam, The Netherlands}

\begin{abstract}
Meteorites contain minerals from Solar System asteroids with different properties (like size, presence of water, core formation). We provide new mid-IR \textcolor{\refrevisions1}{transmission} spectra of powdered meteorites to obtain templates of how mid-IR spectra of asteroidal debris would look like. This is essential for interpreting mid-IR spectra of past and future space observatories, like the James Webb Space Telescope. First we present new \textcolor{\refrevisions1}{transmission} spectra of powdered ordinary chondrite, pallasite and HED meteorites and then we combine them with already available \textcolor{\refrevisions1}{transmission} spectra of chondrites in the literature, giving a total set of 64 \textcolor{\refrevisions1}{transmission} spectra. In detail we study the spectral features of minerals in these spectra to obtain measurables used to spectroscopically distinguish between meteorite groups. Being able to differentiate between dust from different meteorite types means we can probe properties of parent bodies, like their size, if they were wet or dry and if they are differentiated (core formation) or not. 

We show that the \textcolor{\refrevisions1}{transmission} spectra of wet and dry chondrites, carbonaceous and ordinary chondrites and achondrite and chondrite meteorites are distinctly different in a way one can distinguish in astronomical mid-IR spectra. Carbonaceous chondrites type$<$3 (aqueously altered) show distinct features of \textcolor{\refrevisions1}{hydrated silicates (hydrosilicates)} compared to the olivine and pyroxene rich ordinary chondrites (dry and equilibrated meteorites). Also the iron concentration of the olivine in carbonaceous chondrites differs from ordinary chondrites, which can be probed by the wavelength peak position of the olivine spectral features. The \textcolor{\refrevisions1}{transmission} spectra of chondrites (not differentiated) are also strongly different from the achondrite HED meteorites (meteorites from differentiated bodies like 4~Vesta), where the latter show much stronger pyroxene signatures. 

The two observables that spectroscopically separate the different meteorites groups (and thus the different types of parent bodies) are the pyroxene-olivine feature strength ratio and the peak shift of the olivine spectral features due to an increase in the iron concentration of the olivine.
\end{abstract}

\begin{keyword}
Meteorites, Mineralogy, Spectroscopy, Planetary formation, Debris disks
\end{keyword}

\maketitle

\section{Introduction}
Dust grains (micron-sized solid-state particles) play an essential role in many astrophysical environments. One of the prime examples is the formation and evolution of planets in proto-planetary disks. Small micron sized dust grains are the building blocks of planets. Proto-planetary disks are formed from interstellar dust and gas, where the lattice structure of the dust is amorphous and its composition is mostly silicate with some amounts of carbonaceous dust \citep{kemper04, min07_ISMgrains}. In proto-planetary disks this dust can be annealed (the lattice structure is made crystalline due to heating) or it can be condensed from the gas \citep{tielens98,gailsedl99, sogawa99}. Crystalline dust formation takes place in the inner parts of the disk. The crystals can subsequently be mixed with the outer parts of the disk (for example see \citealt{gail04}). 

   \begin{figure*}
   \centering
   \includegraphics[width=0.95 \hsize]{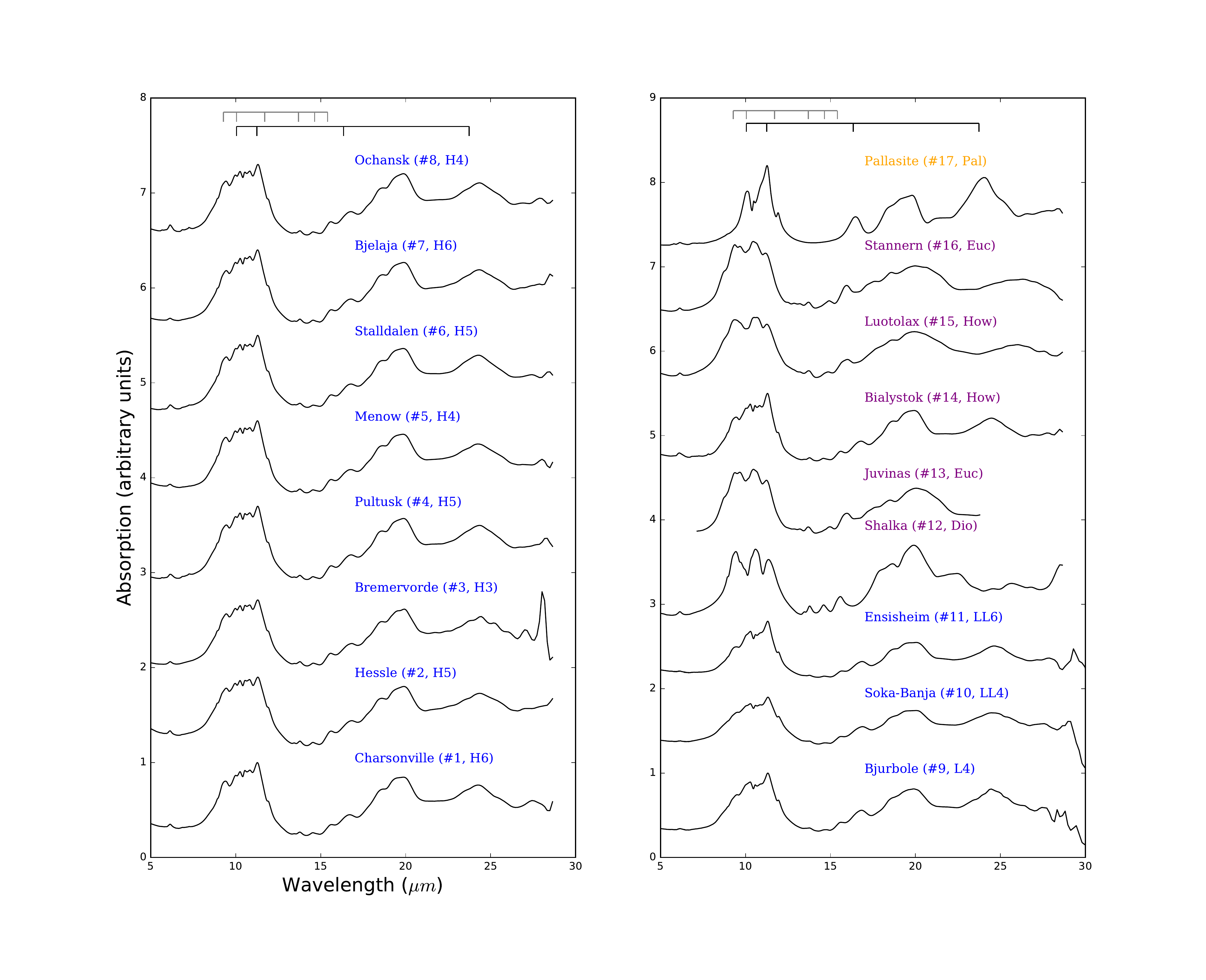}
      \caption{Absorbance infrared transmission spectra of the sample of meteorites presented in this work (see Tab. \ref{tab_sample}). The spectra are scaled and shifted. The ordinary chondrite names are shown in blue, the HED in purple and the pallasite in orange. The black and grey "forks" at the top indicate the olivine and pyroxene feature peak positions (respectively) of the features discussed in this work. The olivine and pyroxene cross-sections are plotted in Fig. \ref{fig_averaged_groups}.}
         \label{fig_sample}
   \end{figure*}
   
The crystalline and amorphous dust in proto-planetary disks can be studied by observing their emission and absorption in the infrared. Much can be learned about the composition, grain size and grain temperature of the crystalline dust in proto-planetary disks by studying and modelling their spectral features \citep{waelkens96, meeus01,spitzer1, sturm13, maaskant14}. The wavelength position, strength and shape of the spectral features show that the crystalline olivine ([Mg,Fe]$_2$SiO$_4$) grains are very magnesium rich (at most the olivine has an Fe/(Mg+Fe) of $\sim$ 0.0-0.03) and crystalline olivine is about 0-10 \% (by mass) of the total dust content in the disk. Besides olivine, pyroxene ([Mg,Fe,Ca]SiO$_3$) is also detected in crystalline form \citep{juhasz10}. Although more difficult to determine, the pyroxene seems to contain more iron than the olivine (for pyroxene \cite{bowey07, sargent09, juhasz10} report Fe/(Mg+Fe) values of 10-25 \%) and in some cases it can have an abundance equal to that of the crystalline olivine. Furthermore, \cite{juhasz10} report that the pyroxene-over-olivine ratio decreases as a function of radius in the disk, which is difficult to explain based on equilibrium considerations \citep{gail04}.

In proto-planetary disks dust grains are used to form planetesimals and eventually planets. When the proto-planetary disk eventually sheds its gas and small dust grains a main-sequence star with a system of planetesimals and planets is left \citep{wyatt08}. During this evolution of the disk the minerals in the planetesimals and planets are influenced by parent-body processes. \textcolor{\refrevisions1}{Depending on the properties and formation process of the parent-body} these processes can be aqueous alteration, equilibration and melting due to heating \citep{hutchison04_book}. Relatively small planetesimals (several to a hundred of kilometers in diameter) can heat up several hundreds of Kelvin due to radiative decay of unstable elements \textcolor{\refrevisions1}{(like \textsuperscript{26}Al)}, during which diffusion between minerals can alter their composition \citep{kessel07}. Larger planetesimals (several hundreds of kilometers and up in diameter) and planets reach temperatures where the minerals melt and the body will differentiate (formation of a core and rocky crust), which has a strong influence on the composition and lattice structure of the minerals. An example of such a planetesimal in our Solar System is the asteroid 4~Vesta \citep{consolmagno77, mccord70}.

The composition of planetesimals can be studied over astronomical distances by observing dynamically active systems where planetesimals have a high probability of colliding and being ground down to micron-sized dust grains \citep{wyatt08}. \textcolor{\refrevisions1}{These micron-sized dust grains can form so called debris disks. It is important to note that the dust in these debris disks does not come from the proto-planetary disk, but consists of a second generation of dust coming from collisions in the system.} Dust grains in debris disks can then be studied by analysing their infrared spectral features. \cite{olofsson12} studied young main-sequence stars with dust relatively close to the star at several hundreds of Kelvin (distances comparable to the asteroid belt). They found olivine with an iron content Fe/(Mg+Fe) $\sim$ 0.2 and Py/(Ol+Py) ratios of 0.0-0.2, which compares well with Solar System asteroids \citep{nakamura11}. The dust produced in a Kuiper-belt like analogue in the system of the main-sequence star $\beta$ Pictoris contains olivine that is very pure Mg-rich (Fe/(Mg+Fe)=0.01) and no pyroxene is found \citep{chen07, devries12}. The debris in the outskirts of $\beta$ Pictoris compares best with cometary olivine found by for example the Stardust mission \citep{zolensky08}.

How minerals change due to parent-body processing can be understood from the meteoritic record of our Solar System. The minerals in meteorites can be studied in detail and they can be linked back to their parent-bodies and their properties. For example ordinary chondrites are linked to relatively small and equilibrated asteroids, some carbonaceous chondrites have been aqueously altered and have been linked to wet relatively small asteroids and the HED (howardite-\textcolor{\refrevisions1}{eucrite}-diogenite) achondrites are an example of achondrites linked to the differentiated asteroid 4~Vesta (for an overview \textcolor{\refrevisions1}{see \citealt{hutchison04_book} and for the Dawn mission \citealt{russell15_dawn})}. Considering the silicate minerals in meteorites, the dominant silicate in carbonaceous chondrites with low metamorphic grade (type 1 and 2) are \textcolor{\refrevisions1}{hydrated silicates (from here on hydrosilicates)}. Silicates in ordinary chondrites have experience little (type 3) to severe (type 4 and up) alteration due to heat which increased their pyroxene abundance and increased the iron content of their olivine and pyroxene minerals (up to Fe/(Mg+Fe) $\sim$ 0.16-0.32 and 0.18-0.26 respectively, \citealt{vanschmus69, brearleyjones98}). The pyroxene content of achondrite meteorites is very high \textcolor{\refrevisions1}{($>$90~vol\%)} and besides pure pyroxene one also observes other phases like plagioclase (a solid-solution between albite (NaAlSi$_{3}$O$_{8}$) and anorthite (CaAl$_{2}$Si$_{2}$O$_{8}$), \citealt{hutchison04_book}). 

\textcolor{\refrevisions1}{Besides detailed compositional information, much work has been done on reflectance and emission measurements (in the optical and near-IR, and both of surfaces and in powdered form, for an overview see for example \citealt{hutchison04_book}). In this work we focus on transmission measurements of powdered meteoritic minerals in de mid-IR (5-25 \mic). We focus on this method because it is directly comparable to observations of dust grains freed by parent body collisions in young planetary systems}. This enables us to make a direct spectroscopic comparison between dust from Solar System asteroids and extra-solar planetesimals and thus learn about the properties and evolution of extra-solar planetesimals. Mid-IR spectra of \textcolor{\refrevisions1}{powdered} meteorites have been done by \cite{morlok10} (also see \citealt{morlok12, morlok14a, morlok14b}), \cite{beck14} and \cite{molster03}. \cite{morlok12} has studied {mid-IR} spectra of several \textcolor{\refrevisions1}{powdered} achondrites and \cite{morlok14b} of several \textcolor{\refrevisions1}{powdered} chondritic meteorites. \cite{beck14} measured the mid-IR spectra of \textcolor{\refrevisions1}{powdered} carbonaceous chondrites. \textcolor{\refrevisions1}{Mid-IR spectra of an interplanetary dust particles (IPD) and {micrometeorites} have been presented in, for example, \cite{sandford85, molster03, bradley99}}. 

The goal of this work is to 1) complement the available sets of mid-IR spectra of \textcolor{\refrevisions1}{powdered} meteorites with missing chondritic and achondritic \textcolor{\refrevisions1}{powdered} meteorite spectra, 2) present an overview of the spectral properties of all \textcolor{\refrevisions1}{powdered} meteoritic mid-IR spectra that are available to date and study their differences and 3) define several quantitative measurements useful for determining the parent-body properties based on the \textcolor{\refrevisions1}{dust and debris in astronomical environments}. The first quantitative measurement on the spectra that we consider in this paper is the relative strengths of the pyroxene and olivine bands as an indication of the pyroxene-over-olivine ratio in the meteorite. The second measurement is the shift in peak position of several olivine and pyroxene spectral features since these shifts are indicative of the iron content of these silicates \citep{koike03}.

The paper is structured in such a way that we first introduce the measurements, sample selection and methods, followed by a presentation of the resulting spectra. Then we explain how we measure the wavelength peak positions of several olivine and pyroxene spectral features as well as the strength of several of these spectral features. This is followed by our results and we end with a discussion and conclusions.

	\begin{table}
		\caption{Meteorite sample of this work}             
		\label{table:1}      
		\centering                          
		\begin{tabular}{l l l}        
		
			Name & 			Type  		& Museum catalogue \\
				&						& number \\
			\hline                        
			\hline                                   
			Charsonville & 		H6  			&NRM\#65:0088\\
			Hessle & 			H5			&NRM\#69:0402\\
			Bremerv\"orde &	H3 			&NRM\#57:0095\\
			Pultusk &			H5  			&NRM\#77:0283\\
			Menow 	&		H4 			&NRM\#LK7805\\
			St\"alldalen &		H5  			&NRM\#76:0347\\
			Bjelaja Zerkov &	H6  			&NRM\#65:0108\\
			Ochansk & 		H4  			&NRM\#93:121\\
			\hline
			Bjurb\"ole &			L4			&NRM\#99:0420 \\
			\hline
			Soko-Banja &		LL4			&NRM\#884239		\\
			Ensisheim &		LL6			&	NRM\#600069  \\
			\hline
			Shalka	 &		diogenite		&NRM\#68:517\\
			Juvinas &			eucrite 		&NRM\#94:0099\\
			Bialystok	 &		eucrite		&NRM\#86:0136\\
			Luotolax	 &		howardit 		&NRM\#69:0664	\\
			Stannern 	&		eucrite 		&NRM\#78:0134\\
			\hline
			Seymchan &		pallasite 		&-\\
			\hline
					
		\label{tab_sample}
		\end{tabular}
	\end{table}

   \begin{figure*}
   \centering
   \includegraphics[width=0.95 \hsize]{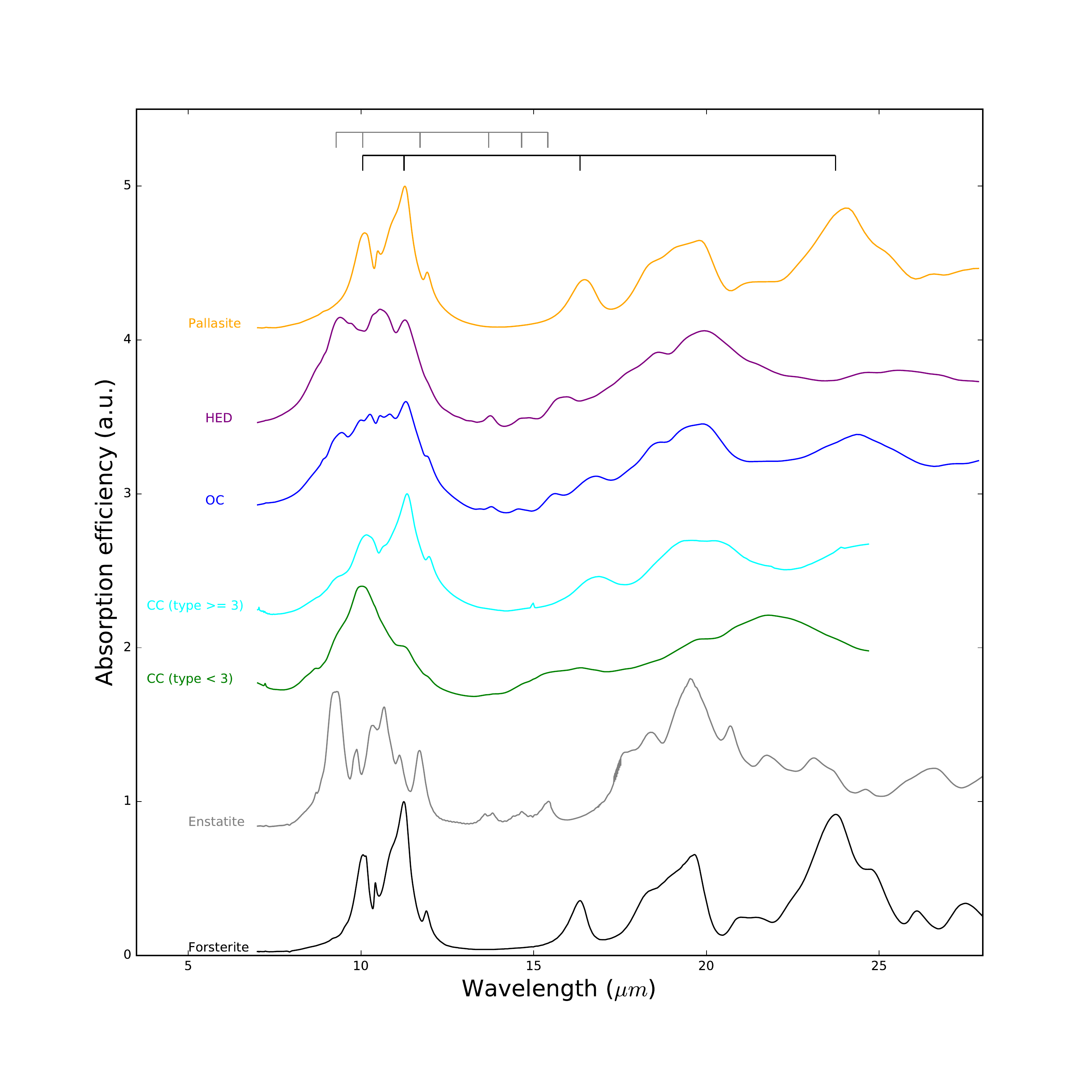}
      \caption{Averaged spectra for the carbonaceous chondrites with metamorphic grade lower than 3 (green), carbonaceous chondrites with type equal and larger than 3 (cyan), ordinary chondrites (blue), HED achondrites (purple) and a pallasite (orange). The "forks" at the top of the plot indicate the wavelength peak positions of pure Mg-rich olivine (black) and pure Mg-rich pyroxene bands (grey) discussed in this work. The opacities of synthetic Mg-rich olivine and Mg-rich pyroxene are shown at the bottom of the plot in black and grey respectively. All spectra are scaled and shifted. The carbonaceous chondrites are from \cite{beck14} and for a comparison with hydrosilicate spectra, please see their work. }
         \label{fig_averaged_groups}
   \end{figure*}

\section{Measurements}

\subsection{Sample selection}
For this study we focussed on meteorites that represent minerals and rocks in medium to large sized planetesimals (ten to hundreds of kilometres in diameter). Measurements of carbonaceous chondrites \citep{beck14} and some ordinary chondrites (OC) \citep{morlok10, morlok12, morlok14a} have been published, but in the literature the OC and HED groups are not completely sampled and the pallasite meteorite group is not measured. These are the meteorites for which we present mid-IR spectra in this work. We collected 13 samples from the meteorite collection of the Swedish Museum of Natural History, Stockholm. Our sample (see table \ref{tab_sample}) consists of 8 H-type OC (ranging from metamorphic grade 3 to 6), one L4 OC and an LL4 and LL6 OC. We also included 6 achondrites to the sample of which five HED and one pallasite. Among the HED we have one howardite, one diogenite and three eucrites. 

Later in this work we will combine the spectra we measured with spectra available in the literature. These are cabonaceous chondrites from \cite{beck14} and some spectra from \cite{morlok12, morlok14a, morlok14b}. We will also compare the meteorite spectra with synthetic laboratory measurements of several minerals. All these measurements and their numbers used in the upcoming plots are listed in tables \ref{fittinginfo1} to \ref{fittinginfo3}.

\subsection{Mid-IR spectra measurement technique}
For the whole sample we obtained Fourier transform infrared (FTIR) transmission spectra in the mid-IR. For the FTIR we needed $\sim$1~mg of sample material. This material was scraped from the meteorite using a diamond top drill and subsequently ground down in an agate mortal for several minutes. Some of the meteorite samples contained metallic iron grains which were removed using a magnet during the grinding process. \textcolor{\refrevisions1}{Metallic iron does not have any spectral features and its absorption in the mid-IR is a smooth continuum (for example see \citealt{ordal88}). Therefor the removal of metallic iron has no impact on our results since we focus on spectral features and not on the continuum absorption.}

The material was then mixed with 200-300 mgr of KBr and pressed into a pellet. A transmission spectrum in the 5-25 \mic range is then obtained using a FTIR spectrometer (Bruker Equinox 55 and Vertex 70). The resulting spectra in absorbance mode are shown in Fig. \ref{fig_sample}. 

Observational studies of dust freed from planetesimals in debris disks sample the bulk of the minerals freed from these planetesimals. Therefor the mid-IR spectra in this study must also be representative of the bulk of the meteoritic material. The drilling method used to free sample material from the meteorite for measurements enabled us to sample the bulk of the meteorite. With the drill we sampled a $\sim$60 mm$^3$ volume of the meteorite. To further ensure we obtain mid-IR spectra of the bulk meteoritic material, we measured the meteorites several times, using different sample locations of the meteorite. In the final measurements no differences between different sample areas of the same meteorite were seen. 

   \begin{figure}
   \centering
      \includegraphics[width=0.9 \hsize]{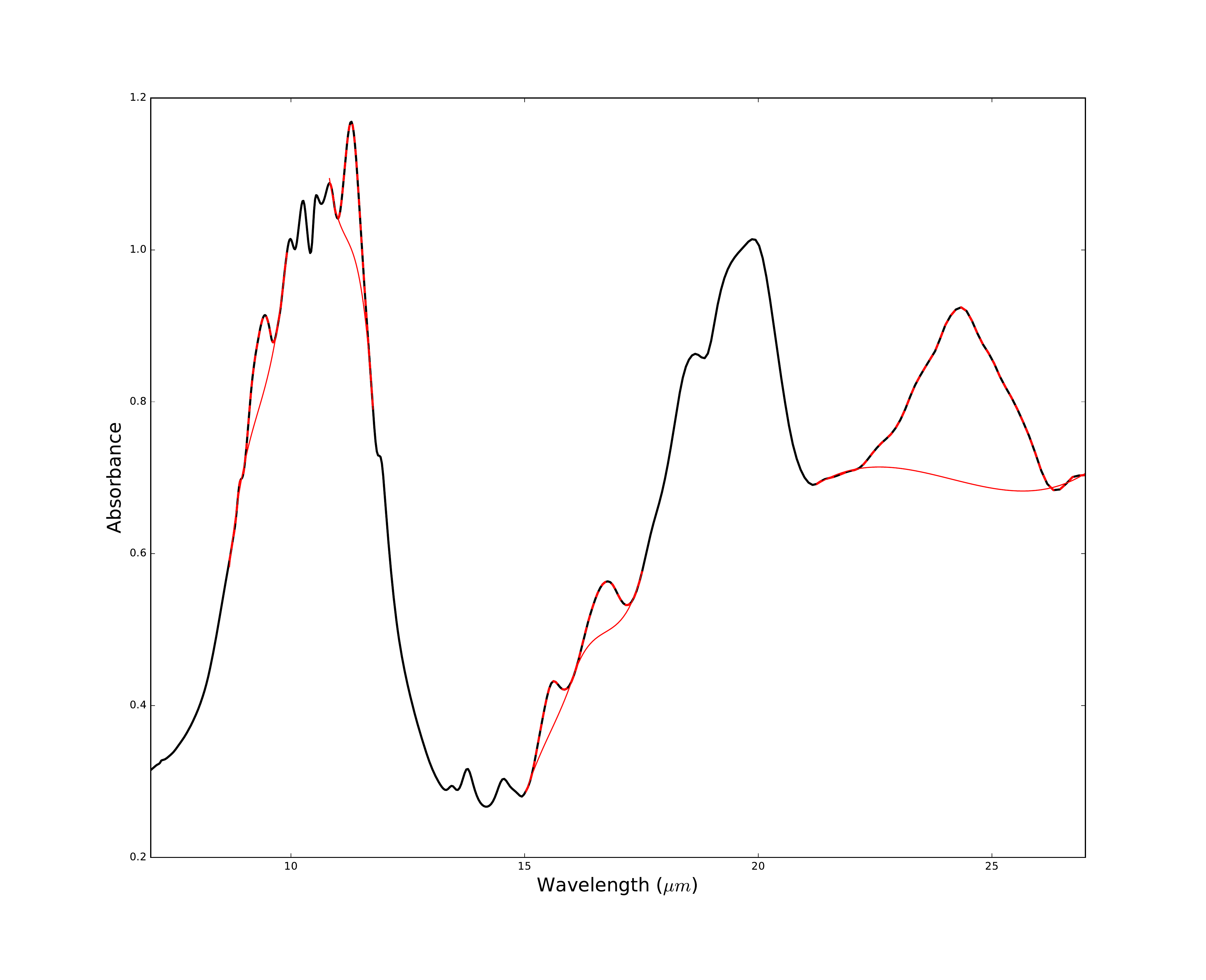}
      \caption{Example of the constructed continua (solid red line) and feature splines (dashed red line) for the case of the spectrum of Bjelaja Zerkov.}
         \label{fig_example_cont}
   \end{figure}

\section{Results}

   \begin{figure*}
   \centering
      \includegraphics[width=0.9 \hsize]{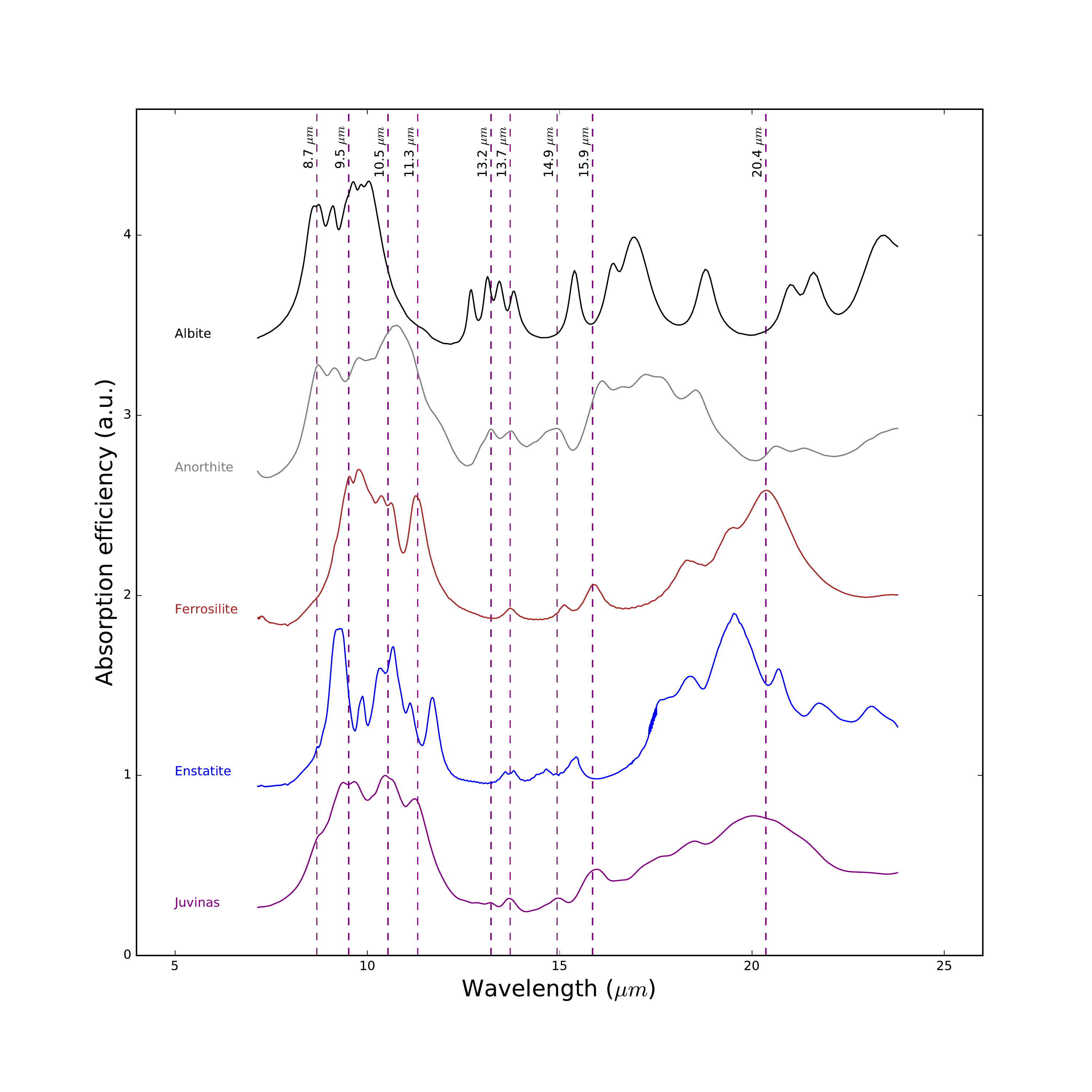}
      \caption{ \textcolor{\refrevisions1}{Comparing the Juvinas (purple) spectrum with Albite (black), Anorthite (grey), ferrosilite (brown) and enstatite (blue). Purple dashed vertical lines indicate features spotted in the Juvinas spectrum. Laboratory spectra of Anorthite (grey) and Albite (black) measured by \cite{salisbury91}. Wavelength peak positions of bands in these spectra close to the bands of olivine and pyroxene are indicated in the plot.} }
         \label{fig_plagio}
   \end{figure*}

In Fig. \ref{fig_sample} we show the measured meteorite mid-IR spectra. From the spectra we can directly observe that those of the OC (H, L and LL types) all look very similar. In the 10-micron region (roughly from 7-13~\mic) we see one large broad feature (due to amorphous silicates) and on top of this we recognise a sharp feature on the blue and the red side. In the 13-17~\mic range we find four weak but distinct features. At longer wavelengths we recognise a strong feature at 19 and 24 \mic. 

The spectra of the HED meteorites look different from the ordinary chondrites, except for Bialystok whose spectrum closely resembles that of the ordinary chondrites. The other HED meteorites show a strong triple structure on top of the 10 \mic complex and they only have three weak bands in the 13-17 \mic region. In the 19-24 \mic region the HED meteorites show a 19~\mic band and Juvinas, Luotolax and Stannern also seem to have a broad feature around 26 \mic, but this is very close to the end of the wavelength range that we were able to measure. 

The pallasite spectrum shows two strong sharp bands in the 10 \mic region and a band at 16 \mic. Similar to the ordinary chondrites the pallasite also shows a 19 and 24 \mic feature.

   \begin{figure*}
   \centering
      \includegraphics[width=0.9 \hsize]{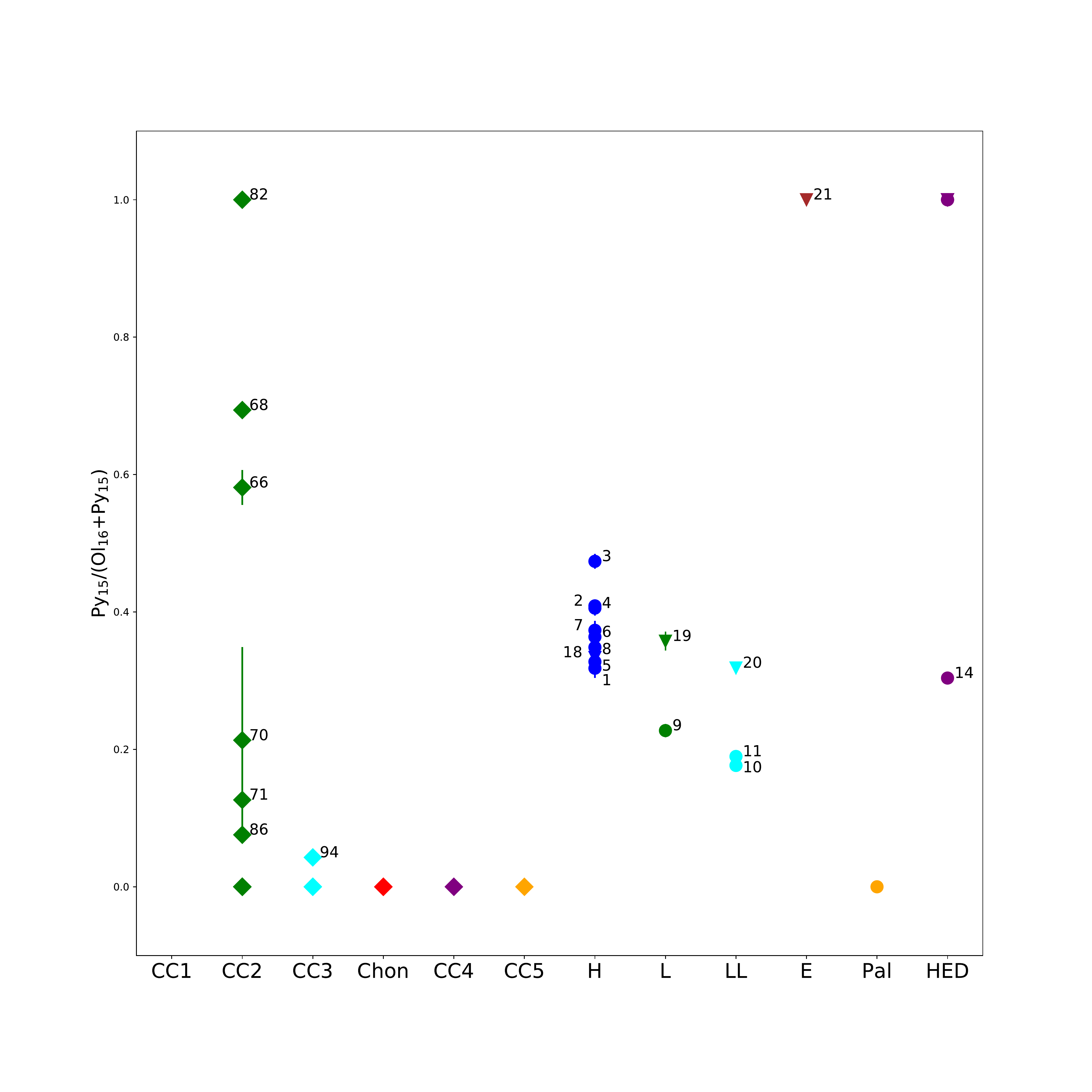}
      \caption{The $\frac{Py}{Py+Ol}$ ratio for the different meteorite groups as determined from the 15 and 16~\mic bands. \textit{CC1}, \textit{CC2}, up to \textit{CC5} stands for carbonaceous chondrites of metamorphic grade 1 and up. \textit{Chon.} stands for chondrules, which are the chondrules of the Allende meteorite (\#88). \textit{H, L} and \textit{LL} stand for ordinary chondrites of that type, \textit{E} stands for enstatite chondrites and \textit{Pal.} for pallasite. For clarity some data points are not numbered when the ratio is either one or zero, the source numbers can be deduced from the numbering of the other sources in their group. The blue, green, cyan, purple and orange dots and triangles are data points for the OC H-type, OC L-type, OC LL-type, HED and pallasite meteorites respectively. The dots are the meteorites presented in this work, while the triangles are meteorites of \cite{morlok12, morlok14a}. The blue, green, cyan, purple and orange diamonds are the type 1, 2, 3, 4, 5 CC meteorites \citep{beck14} and the red diamonds are the chondrules of the Allende meteorite \citep{morlok14a}}
         \label{fig_ratio}
   \end{figure*}

In Fig. \ref{fig_averaged_groups} we show averaged spectra of the OC and HED group meteorites in order to study their general spectral properties. In the figure we compare the meteorite spectra to laboratory measurements of forsterite (Mg-rich olivine, \citealt{koike03}) and enstatite (Mg-rich pyroxene, \citealt{chihara02}), which allow us to identify all the spectral features. The pure Mg-rich olivine and pure Mg-rich pyroxene laboratory measurements shown in Fig. \ref{fig_averaged_groups} are measured with the same method as the meteorite spectra presented in this work. 

In Fig \ref{fig_sample} and Fig. \ref{fig_averaged_groups} we indicated the positions of the most notable features in the spectra of Mg-rich olivine and Mg-rich pyroxene as "forks". In the spectra of ordinary chondrites we can now identify the sharp features on the blue ($\sim$9 \mic) and red ($\sim$11 \mic) flank of the 10~\mic complex as pyroxene and olivine respectively. Of the four bands in the 13-17~\mic region of the ordinary chondrites the first three are from pyroxene and the one at 16~\mic is from olivine. The 19~\mic band in the ordinary chondrite spectra is a combination of olivine and pyroxene. The strong and broad band at $\sim$24~\mic is from olivine. The features in the ordinary chondrite spectra confirm that these meteorites contain significant amounts of olivine and pyroxene. When we closely inspect the wavelength position of the features in the 13-17 \mic range and of the 24 \mic band, we see that the features in ordinary chondrites are red shifted compared to those of Mg-rich olivine and Mg-rich pyroxene. We know that most features of olivine and pyroxene shift to longer wavelengths as the iron content of the mineral increases (\citealt{koike03, chihara02}). This shift is most notable for features in the 15-30~\mic range compared to the features in the 10-micron complex. Thus we can already see that the olivine and pyroxene in the ordinary chondrites contain more iron than the Mg-rich olivine and pyroxene plotted in Fig. \ref{fig_sample} and Fig. \ref{fig_averaged_groups}.

Comparing the HED spectra to those of Mg-rich olivine and Mg-rich pyroxene we see that all (except Bialystok) HED spectra are very similar to that of pyroxene. They all have the three peaks in the 10~\mic range as well as the three weak bands in the 13-17~\mic region. In the HED spectra we do not recognise any olivine, since we see no bands at 16 or 24~\mic. When we inspect the wavelength positions of the features in the 13-17~\mic range we see that the bands in the HED spectra are red shifted compared to Mg-rich pyroxene. This indicates that the pyroxene in the HED could contain significant amounts of iron. Bialystok is an exceptional HED in our sample since its mid-IR spectra looks similar to the ordinary chondrites and thus shows features of both olivine and pyroxene. 

\textcolor{\refrevisions1}{HED meteorites (and especially eucrites) are known to contain plagioclase minerals. To investigate if plagioclase features are visible in the measured spectra, we compare the spectrum of Juvinas to that of albite (NaAlSi$_{3}$O$_{8}$) and anorthite (CaAl$_{2}$Si$_{2}$O$_{8}$) in Fig. \ref{fig_plagio} \citep{salisbury91}. We also show the spectrum of the Mg and Fe rich end members of pyroxene. Fig. \ref{fig_plagio} shows that the spectrum of Juvinas is dominated by that of Fe-rich pyroxenes. At 8.7 and 13.2 \mic we see distinct, although weak, bands of anorthite. At 9.5, 13.7 and 15 \mic we likely have a blend of anorthite and pyroxene bands.}

The last spectrum we measured is the pallasite spectrum and we can directly recognise its spectrum as olivine without any signs of pyroxene or other minerals. Comparing the pallasite spectrum to Mg-rich olivine we see that the 16 and 24~\mic bands of the pallasite are red shifted and thus the olivine will contain some amounts of Fe. Interesting about the peak positions of the bands in the pallasite spectrum is that the amount by which the features are shifted is less than the shift for the ordinary chondrites. This hints at a lower iron content of the olivine in the pallasite compared to the ordinary chondrites.

Fig. \ref{fig_averaged_groups} shows the averaged spectra of the meteorites measured in this work, but also includes the averaged spectra of carbonaceous chondrites (CC) measured by \cite{beck14}. We have subdivided and averaged the CC meteorites into two groups: one containing the CC meteorites with metamorphic grades 1 and 2 and one group with grades of 3, 4 and 5. The type 1 and 2 CC meteorites have been aqueously altered, where 3, 4 and 5 have not and 4 and 5 have equilibrated due to heating in their parent body. The averaged spectrum of the CC type$<$3 meteorites is dominated by hydrosilicates, which shows broad bands at 10, 15 and 22 \mic (see \citealt{beck14}). The type$<$3 CC meteorites show very weak but still distinguishable olivine bands at 11~\mic and for some spectra at 16~\mic (not recognisable in the averaged spectrum, see \citealt{beck14}). In contrast, the CC type$\geq$3 spectra are dominated by olivine and in some cases minor bands of pyroxene can be detected (for example at 9~\mic). 

Now that we have a full overview of the spectra of different meteorite groups in Fig. \ref{fig_averaged_groups}, it is interesting that these five groups look very different, meaning that we can spectroscopically recognise them. In the next section we will study the wavelength position of the olivine and pyroxene features in the meteorite spectra in more detail.

\section{Quantifying the spectral differences between the meteorite groups}
In this section we will study the spectral difference between the meteoritic groups in a quantitative way. We will get a measure of the ratio of pyroxene to olivine in the meteorites by calculating the $\frac{Py}{Py+Ol}$ ratio. For $Py$ and $Ol$ we take the integrated strength of the 15~\mic and 16~\mic bands for pyroxene and olivine respectively. In choosing these features it was important to pick two features that are close together in order to minimise any wavelength and temperature effects. For example in astronomical spectra the emission of the features is dependent on temperature in addition to abundance, but the temperature effect is divided out in the ratio when the features are close together in wavelength. Another set of features to consider for the pyroxene over olivine ratio are the 9~\mic and 11~\mic bands. The difficulty with the 11~\mic band of olivine is that it overlaps with a pyroxene band at the same wavelength. Also the strength and wavelength position of these two features are very sensitive to grain size (\citealt{devries15}), making them a less ideal choice.

We study the shift in the olivine and pyroxene features by measuring the peak wavelength positions for two features for each mineral. For olivine we will focus on the features at 16 and 24~\mic. Both these features are nicely isolated (do not overlap with pyroxene features) and can therefor be easily measured. These two features also show prominent shifts due to increasing iron content in their lattice \citep{koike03}. For pyroxene we will measure the feature at 9 and 15~\mic for similar reasons.

\subsection{Method of measuring spectral feature properties}

In order to measure the wavelength peak positions \textcolor{\textNew}{and the strength} of features we construct a continuum under each feature. For this we define a region to the left and right of the feature and fit a spline through the continuum points (the continua points used are listed for all objects in Tab. \ref{tab_cont_points} in the appendix). The spline fit is done using \textit{interpolate.splrep} from the \textit{Scipy Python} package. In order to measure the peak position the continuum subtracted feature is then fitted with a spline to smooth the spectrum. The peak wavelength position is then found by finding the root of the derivative of the spline of the continuum subtracted feature (using the \textit{Scipy} function \textit{scipy.optimize.fsolve}). For the ratio $\frac{Py}{Py+Ol}$ we measure the integrated strength of the features by integrating over the continuum subtracted transmission efficiency. An example of the continua and splines we obtain are given in Fig.~\ref{fig_example_cont}. 

In order to estimate the errors in the derived peak positions and feature strengths, we use a Monte Carlo approach. We calculate these values for features at 1000 iterations and at every iteration we vary our choice of the continuum points left and right of the feature. We randomly vary the boundaries of the continua we choose left and right of the feature at every iteration by 15~\% using a uniform distribution (using the \textit{Python} standard \textit{random.uniform}). The peak position and strength of the feature is then estimated by taking the average over all iteration. The error we state in this paper is the standard deviation of the peak position or strength of the feature over all iterations. The resulting values we measured for all spectra mentioned in this paper are listed in Tab. \ref{fittinginfo1}, \ref{fittinginfo2} and \ref{fittinginfo3}.

\subsection{Pyroxene over olivine ratio}
\textcolor{\textNew}{In Fig. \ref{fig_ratio} we show the $\frac{Py}{Py+Ol}$ ratio for every meteorite group discussed in this paper. It shows that the OC consistently have a high $\frac{Py}{Py+Ol}$ ratio, where the H-type have the highest at 0.4 and it drops of to 0.2 for the L and LL chondrites. The HED (except Bialystok \#14, which is similar to the OC) and the enstatite chondrite have ratios of 1.0. All the CC have a ratio of 0.0, except for the type-2 carbonaceous chondrites, which show a large spread. The CC type 2 that show the spread are CM and CR carbonaceous chondrites. For the pallasite meteorite we measured a ratio of 0.0.
}

   \begin{figure*}
   \centering
      \includegraphics[width=0.9 \hsize]{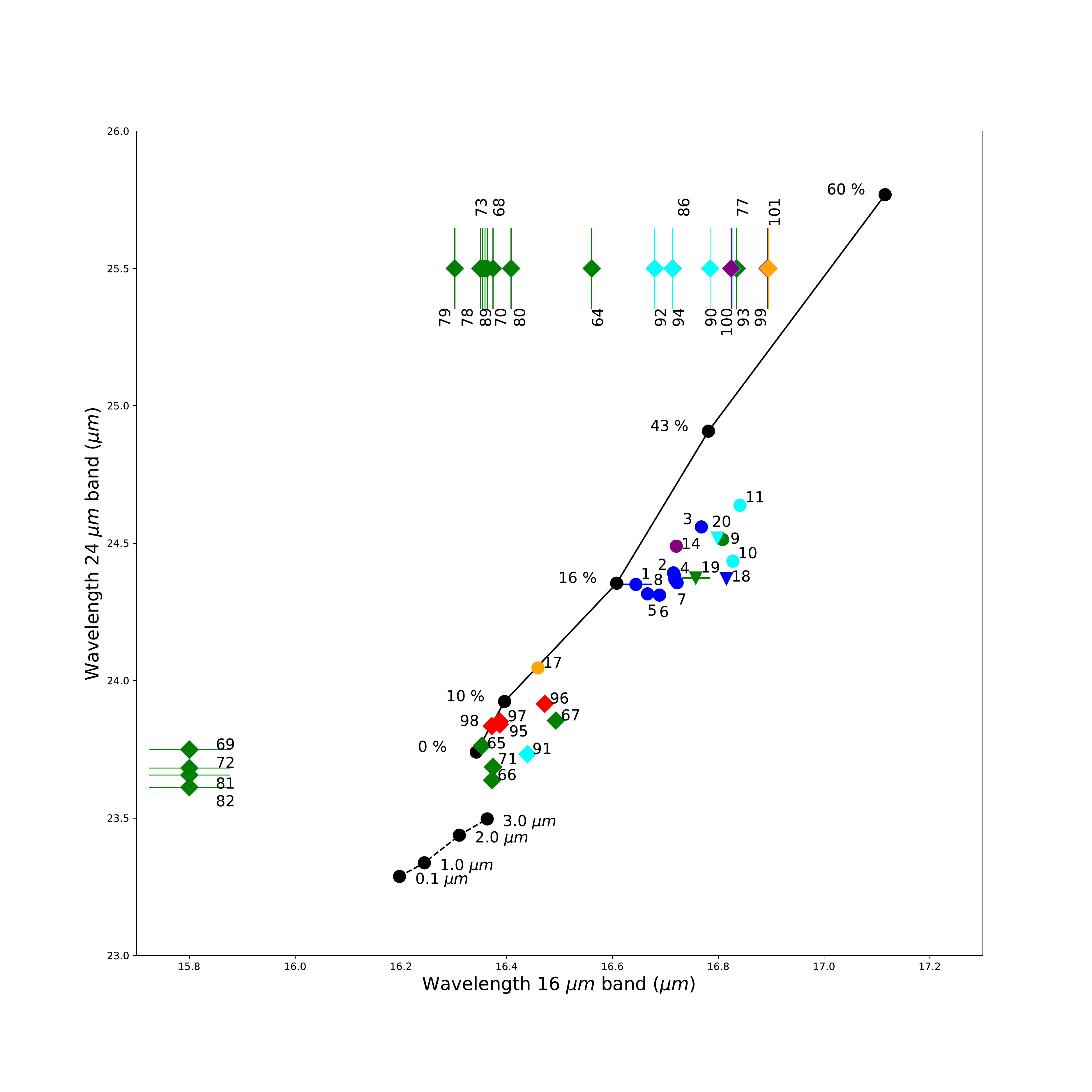}
      \caption{Wavelength peak positions of the 16 and 24 \mic bands of olivine for meteorites as well as synthetic crystals. The black dots connected by a black solid line are the wavelength positions for olivine with different concentrations of Fe as measured by \cite{koike03}. The percentage next to the dots indicates the amount of iron in the olivine. The black dots connected with a dashed line are wavelength positions for opacities of pure forsterite, calculated from optical constant measured by \cite{servoin73}. The calculations from optical constants to opacities are done for different grain sizes, indicated next to the data points (further more the particle shape is according to the Gaussian Random Field method, see \cite{min08}). The blue, green, cyan, purple and orange dots and triangles are wavelength positions for the OC H-type, OC L-type, OC LL-type, HED and pallasite meteorites respectively. The dots are the meteorites presented in this work, while the triangles are meteorites of \cite{morlok12, morlok14a}. The blue, green, cyan, purple and orange diamonds are the type 1, 2, 3, 4, 5 CC meteorites \citep{beck14} and the red diamonds are the chondrules of the Allende meteorite \citep{morlok14a}. Spectra for which only one of the two features could be measured the measurement is displayed as a symbol with a large bar at 15.8~\mic or 25.5~\mic.
}
         \label{fig_16_24_carbon}
   \end{figure*}

\subsection{Peak positions in synthetic mineral spectra with different iron content and grain sizes}
Before reporting on the feature wavelength positions of the meteoritic spectra, we first show how the peak wavelength positions of the olivine and pyroxene bands change as a function of the Fe content of the minerals and as a function of grain size. In Fig. \ref{fig_16_24_carbon} we show the wavelength peak positions of the 16 and 24~\mic bands of olivine with different Fe contents. We use the data of \cite{koike03} because their laboratory measurements have been done in the same way as our meteorite measurements (powdered grains embedded in a KBr pellet). The black dots connected with a solid line in Fig. \ref{fig_16_24_carbon} show how both the 16 and 24~\mic bands shift to longer wavelengths as the Fe content in the olivine increases. 

The black dots connected with a dashed curve in Fig. \ref{fig_16_24_carbon} show peak positions for grains with sizes of 0.1 to 3.0 \mic. These peak positions are calculated from the opacities (cross sections) of grains with different sizes. These opacities were calculated from the optical constants of pure forsterite (measurements done by \citealt{servoin73}). For the calculations of optical constants to opacities we used a Gaussian Random Field (GRF) model for the grain shape (see \citealt{min08}). This state-of-the-art model for grains approaches the best that of irregular grains. Fig. \ref{fig_16_24_carbon} shows that as a functions of grain size the 16 and 24~\mic bands shift to longer wavelengths. When the grains become larger than 3 \mic, the mid-IR bands start to weaken so much that they cannot be measured. The peak positions of the GRF opacities do not connect to those of the \cite{koike03} measurement of pure forsterite. This is likely due to effects introduced by differences in the grain shape and/or measuring technique between the GRF particles and the \cite{koike03} measurements.

In Fig. \ref{fig_9_15_carbon} we show a similar plot as Fig. \ref{fig_16_24_carbon}, but now for the pyroxene bands at 9 and 15~\mic. The solid black curve shows pyroxene for different Fe contents. We use the measurements of \cite{chihara02}, who have measured pyroxene with different Fe contents in the same way as our meteorite spectra (powdered grains embedded in KBr pellets). Fig. \ref{fig_9_15_carbon} shows how both the 9 and 15~\mic bands shift to longer wavelengths when the iron content increases in the mineral. The dashed curve shows peak positions for different grain sizes of opacities calculated with GRF particles of optical constants of enstatite \citep{jager98}. The dashed curve shows that the 9~\mic band is sensitive to grain size, while the 15~\mic band is not. 

\subsection{Peak positions in meteoritic spectra}

   \begin{figure*}
   \centering
      \includegraphics[width=0.9 \hsize]{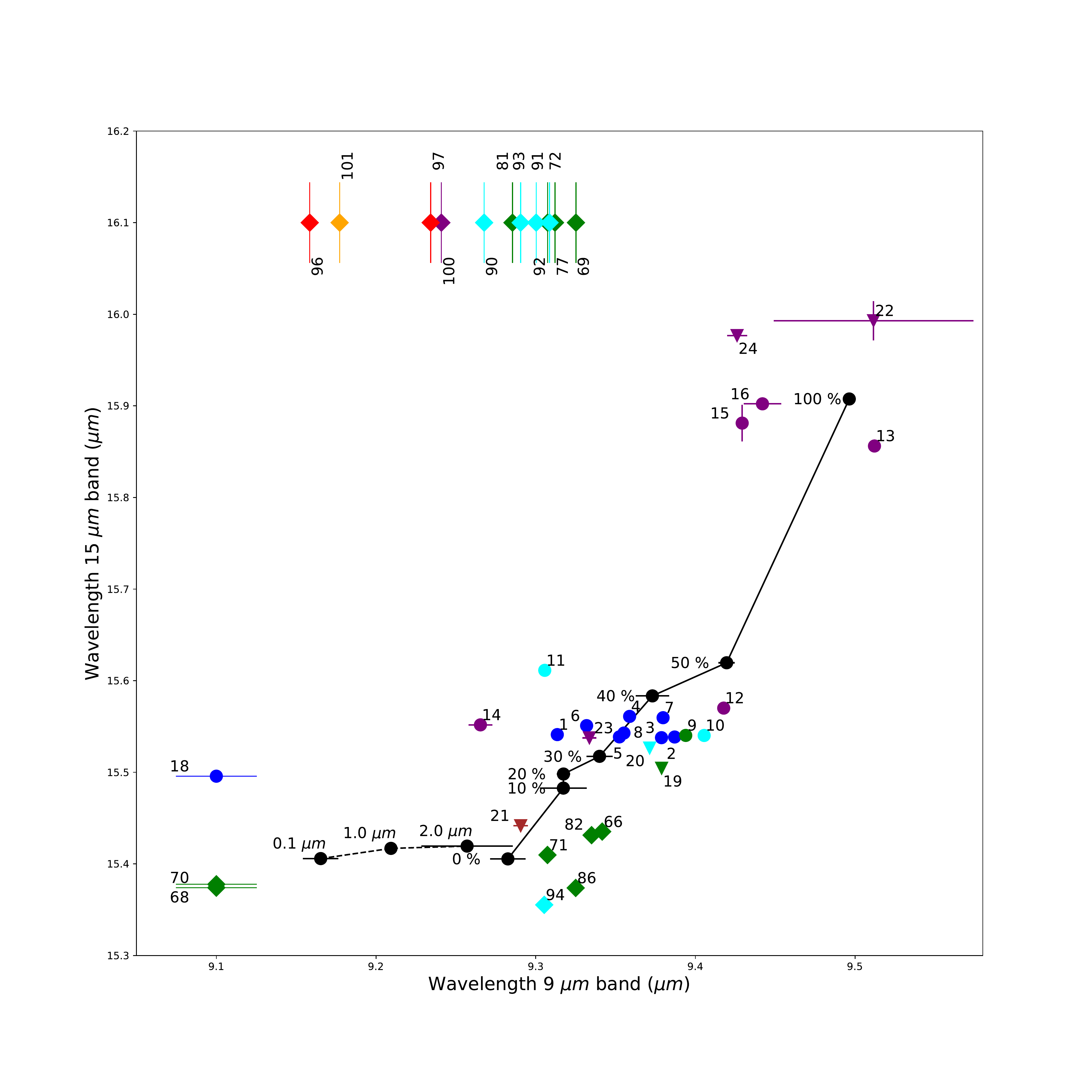}
      \caption{The same plot as Fig. \ref{fig_16_24_carbon} but for the 9 and 15 \mic band of pyroxene.  }
         \label{fig_9_15_carbon}
   \end{figure*}

The peak positions of the olivine bands in the meteorites presented in this work (see Tab. \ref{tab_sample}) are shown in Fig. \ref{fig_16_24_carbon}. Most of the HED meteorites do not show any olivine bands, except for Bialystok (\#14). All the ordinary chondrites show olivine bands with very similar peak positions. The peak position of the pallasite Seymchan (\#17) is also shown in Fig. \ref{fig_16_24_carbon}.  From Fig.~\ref{fig_16_24_carbon} we see that the peak positions of the olivine bands of the ordinary chondrites compares best with those of olivine with 16-40 \% iron. Among the ordinary chondrites the H-type spectra have the shortest wavelength peak positions compared to the L-type and LL-type, although the differences are small. The pallasite Seymchan (\#17) has the bluest peak positions of its olivine bands among the meteorites presented in this work and compare best with olivine with 10-16~\% iron.

The peak positions of the pyroxene bands in the meteorite sample are shown in Fig. \ref{fig_9_15_carbon}. The ordinary chondrites show a spread in the peak positions of the 9 \mic band, but very little variation in the 15 \mic band peak position. It is difficult to say if the spread in the 9~\mic band position of the OCs is due to difference in grain size or iron concentration. But we can conclude that the peak position of the OCs compares best with pyroxene that has a 10-50~\% concentration of iron. In the same way we saw this for the olivine bands of Bialystok (HED, \#14), the pyroxene bands of Bialystok (\#14) also have peak positions very similar to those of the OCs. The pyroxene bands of the HED Shalka (\#12) are also very similar to those of the OCs. The other HED meteorites have pyroxene bands very different from the OCs. Juvinas (\#13), Luotolax (\#15) and Stannern (\#16) all have strongly red shifted 15~\mic bands compared to pure enstatite. Among these three Juvinas (\#13) and Stannern (\#16) also have 9~\mic bands that are red shifted and their peak position compares well with pyroxene with very high concentrations of iron (up to 100\%, called ferrosilite). Luotolax (\#15) has a 15~\mic band that compares well with bands of ferrosilite, but its 9~\mic band is at too short wavelengths compared to those of synthetic ferrosilite.  

In Fig. \ref{fig_16_24_carbon} and \ref{fig_9_15_carbon} we also added OC and HED meteorites available from other sources. We show peak positions of the OC meteorites Cenniceros (\#18), Barratta (\#19) and Parnallee (\#20) measured by \cite{morlok14a}. The peak positions of these OC meteorite spectra are similar to the OC meteorites measured for this work. In Fig.~\ref{fig_9_15_carbon} we also show the pyroxene peak positions of the HED meteorites measured by \cite{morlok12} (\#22 and 24). The peak positions of these HED meteorites are closest to the bands of ferrosilite, like the HED meteorites Juvinas (\#13) and Stannern (\#16) from the sample presented in this work. The \cite{morlok12} sample also contained an enstatite chondrite, Indarch (\#21). As expected this meteorite does not show olivine bands and its pyroxene bands compare best with pyroxene with low iron concentrations.

We also include the measurements of mid-IR spectra of carbonaceous chondrites. We collected carbonaceous chondrites measured by \cite{beck14} and mid-IR spectra of the chondrules from Allende (type CV, metamorphic grade 3, \citealt{morlok14a}). \cite{beck14} measured carbonaceous chondrites of metamorphic grade 1 up to 5. The mid-IR spectra of different metamorphic grades look very different (see Fig. \ref{fig_averaged_groups}). Where grade 1 is dominated by features of hydrosilicates and the spectra become increasingly more silicate-like when going to higher metamorphic grades, where type 4 and 5 can be completely dominated by olivine and some pyroxene.

Fig. \ref{fig_16_24_carbon} shows the olivine bands of the carbonaceous chondrites of \cite{beck14}. Some of the spectra have both measurements of the 16 and 24~\mic band, but most only have a 16~\mic band because the measured spectra did not extend into the 24~\mic range. Striking is that most type 2 and 3 carbonaceous chondrites have 16 and 24~\mic bands that compare best with laboratory measurements of 0-10\% iron, while higher metamorphic grades have 16~\mic bands that compare best with 15-50\% iron concentrations. The chondrules of the Allende meteorite are also shown in Fig.~\ref{fig_16_24_carbon} with red diamonds and they compare best with laboratory measurements with $\sim$0-10\% iron. This in contrast to the bulk measurement of Allende (\#88) of \cite{beck14}, which has a 16~\mic band that compares best with 43\% iron concentrations.

In Fig. \ref{fig_9_15_carbon} we show the pyroxene bands of the carbonaceous chondrites. Five meteorite spectra have both the 9 and 15~\mic bands and these are of metamorphic type 2 and 3 and these bands indicate a low iron concentration for the pyroxene. There are quite a few carbonaceous chondrites that only have a 9~\mic band and no 15~\mic band. These~9 \mic bands are weak, which might explain why the 15~\mic band is not detected for these meteorites. Difficult to explain is that some of the meteorites (type 4 and 5 and Allende chondrules) have 9~\mic bands that are blue shifted compared to the pure enstatite (0\% iron) bands.

\section{Discussion}

	\begin{table*}
		\caption{Meteorite parent body properties and the spectroscopic observables of their debris}             
		\label{table:1}      
		\centering                          
		\begin{tabular}{l l l l l }        
			\hline
			Group						& Parent body	type 				& Spectrum				& Peak shifts indicate		&  Feature ratio  \\
										& 			 				& 						& Fe/(Mg+Fe)				&  $\frac{Py}{Py+Ol}$  \\
			\hline
			\hline
			
			Carbonaceous chondrites  		& Aqueous alteration  				& Hydro-silicate bands  		& Ol: 0.00-0.10				& 0.0 or a			\\
			type 1 \& 2	 				& Relatively small 					& Possibly weak Ol and Py  	& Py: 0.00-0.10				& large spread						\\
										& ($\sim$10-100 km)					&						&						&							\\
			\hline
			Carbonaceous chondrites  		& Equilibration		  				& Olivine					& Ol: 0.15-0.40				& 0.0							\\
			type $\geq$ 3	 				& Relatively small 					& Weak pyroxene			&						& 				  			\\
										& ($\sim$10-100 km)					&						&						&							\\
			\hline
			Ordinary chondrites  				& Equilibration		  				& Olivine \& pyroxene		& Ol: 0.15-0.40				& 0.2-0.4		 				\\
						 				& Relatively small 					& 						& Py: 0.10-0.50				&							\\
										& ($\sim$10-100 km)					&						&						&							\\
			\hline
			HED			  				& Differentiation \&	  				& Dominated by			& Py: High 				& 1.0							\\
						 				& Igneous processes	 	 			& pyroxene-like				& Can contain other 			&							\\
				 						& Relatively large 					&  appearance				& phases then 				&							\\
										& ($\sim$250 km)					&						& Mg-Fe-pyroxene			&							\\
			\hline
			Pallasites						& Impact-generated 					& Pure olivine				& Ol: 0.10-0.16				& 0.0			 				\\
				 						& mixture or from 			 	 	& 						&						&							\\
										& crust-core boundary				&						&						&							\\
			\hline
		\label{tab_meteorite_spectra_summary}
		\end{tabular}
	\end{table*}

We measured and investigated the mid-IR spectra of several groups of meteorites. In detail we studied the peak positions of the 16 and 24~\mic bands of olivine, the 9 and 15~\mic bands of pyroxene and the $\frac{Py}{Py+Ol}$ ratio. The peak positions of the spectral bands are dependent on the composition of the silicates. Through the use of the shift in the 16 and 24~\mic bands of olivine we showed that the olivine iron content of ordinary chondrites ranges between 15-40 \% (see Fig. \ref{fig_16_24_carbon}). Here the H-type OC seem to have slightly lower olivine iron concentrations compared with the few L and LL-type OC we measured. We do not see much difference between different metamorphic grades. Similar for the 9 and 15~\mic bands of pyroxene we showed that the OC have pyroxene iron concentrations between 10-50 \% (see Fig.~\ref{fig_9_15_carbon}). Both the olivine and pyroxene compositions are consistent with detailed compositional studies which show 16-20~\%, 22-26~\% and 27-32~\% iron concentrations in the olivine for H, L and LL-type OC respectively and 18-26~\% iron concentrations in the pyroxene for H, L, LL-type meteorites \citep{vanschmus69, brearleyjones98}. Based on the $\frac{Py}{Py+Ol}$ ratio we showed that the pyroxene and olivine content of OCs gives a ratio of 0.4 down to 0.2.

For many of the carbonaceous chondrites the spectra of \cite{beck14} did not reach long enough wavelengths to measure the 24 \mic band of olivine, also many carbonaceous chondrites have such low pyroxene compositions that the 15~\mic band could not be detected. The $\frac{Py}{Py+Ol}$ ratio showed that most carbonaceous chondrites are pyroxene poor, except for some CM and CR types that show a large spread of ratios. From the band positions in Fig.~\ref{fig_16_24_carbon} we could see that the aqueously altered and un-equilibrated (metamorphic grade 2 or 3) carbonaceous chondrites have a very low (0-10 \% based on the band positions) iron concentration in the olivine. This is consistent with composition measurements, which indicate olivine iron concentrations of 0-5 \% for these un-equilibrated carbonaceous chondrites \citep{rubin88, kallemeyn94, bischoff93}. From the 16~\mic olivine bands for equilibrated carbonaceous chondrites (metamorphic grades 4 and 5) we find higher olivine iron concentrations, which are of the order of the concentrations in ordinary chondrites. Such higher iron concentrations in the olivine are confirmed by laboratory measurements reviewed in \cite{hutchison04_book}.

The HED meteorites in our sample consist of three groups, the eucrites, diogenites and howardites. These meteorites are rocks from a body other than Earth, the Moon or Mars, which experienced basaltic vulcanism and igneous fractionation. The most likely parent body for these meteorites is 4~Vesta \citep{takeda97, russell15_dawn}. HED meteorites are known to be composed mostly of pyroxene and/or plagioclase and minor amounts of other minerals \citep{hutchison04_book}. Here plagioclase is a solid-solution between albite (NaAlSi$_{3}$O$_{8}$) and anorthite (CaAl$_{2}$Si$_{2}$O$_{8}$). Of the HED meteorites, diogenites are rich in ($>$90 vol\%) pyroxene, while eucrites are rich in plagioclase ($>$90~vol\%). Diogenite's pyroxene composition ranges within Wo$_{1-4}$En$_{77-67}$Fs$_{20-29}$ \citep{fowler94} (where Wo stands for wollastonite, which is a calcium-rich pyroxene end-member). The plagioclase in eucrites can have compositions of for example An$_{93-80}$ in Juvinas (\#13) and An$_{89-72}$ in Stannern (\#16) \citep{takeda97}. Howardites are surface breccias composed of diogenite, eucrite and sometimes carbonaceous chondrite debris. 

\textcolor{\refrevisions1}{The eucrites, diogenites and howardites have distinct mid-IR spectral characteristics. For example the eucrites (Juvinas \#13, Stannern \#16 and \#22 from \citealt{morlok12}) all have pyroxene bands that compare best with Fe-rich pyroxenes. In Fig. \ref{fig_plagio} we compared the spectrum of the eucrite Juvinas to that of anorthite and albite, but we did not find strong features in the Juvinas spectrum that indicated any plagioclase. This is likely due to a difference in cross section values of pyroxene and plagioclase. The diogenites (Shalka \#12 and \#23 from \citealt{morlok12}) in our sample compare well with pyroxene with iron concentrations of 20-40 \%, which is similar to the range found in laboratory studies of Fs$_{20-29}$ \citep{fowler94}. The pyroxene-rich nature of the HED group is reflected in the $\frac{Py}{Py+Ol}$ ratio, which is $\sim$1.0 for all of them.}

The howardite meteorites in our sample are all very different, which likely reflects their brecciated nature. For example some (\#15 and \#24) have pyroxene bands that are closest to 100\% ferrosilite (like eucrites). Bialystok (\#14) has pyroxene bands that compare well with diogenites. Interesting about Bialystok is that its spectrum (see Fig. \ref{fig_sample}) looks very similar to the ordinary chondrites. Bialystok is also the only HED meteorites for which we detected olivine bands. Besides the HED achondrite meteorites we also have an achondrite pallasite meteorite in our sample (Seymchan, \#17). The pallasite spectrum confirms the fact that pallasites contain only the silicate olivine ($\frac{Py}{Py+Ol}$ ratio is 0.0). The pallasite's mid-IR bands compare best with olivine with an iron concentration of 10-16\%. 

The main result of this work is that the mid-IR spectra of debris from aqueously altered, pristine, equilibrated and differentiated parent bodies show distinct and observable differences. Or in other words, we can observe from the spectra of the debris if its parent body was wet or dry and if it was pristine, small or large. When going from an aqueously altered (small and wet) to a pristine (small), equilibrated (dry and small) and then a differentiated parent body (large), the debris of such a body will show a spectrum with a decrease in the hydrosilicate content, an increase in the pyroxene/olivine ratio and a red-shifting of the olivine bands due to an increase of iron in the olivine. These spectroscopic differences and their link to the properties of their parent bodies are summarised in Tab. \ref{tab_meteorite_spectra_summary}.

Small (micron-sized) grains of olivine and pyroxene have been observed in many astronomical environments in both crystalline and amorphous form. Examples of such environments are disks around pre-main-sequence stars \citep{waelkens96, meeus01,spitzer1}, main-sequence stars \citep{chen07, olofsson12}, comets \citep{wooden02}, post-main-sequence stars \citep{waters96, syl99, mol02_1} and active galaxies \citep{kemper07, spoon06}. Using the results of many studies of planet forming disks we now briefly discuss how the olivine and pyroxene in the meteorites compares with the grains found in these environments. When interpreting these studies one has to consider the techniques and laboratory measurements used to obtain the dust cross sections. The common techniques to obtain cross sections of dust grains are 1) transmission spectroscopy of in KBr embedded samples, 2) model calculations of cross-sections from optical constants (obtained from reflection measurements) and, 3) transmission spectroscopy of grains suspended in a gas (aerosol measurements). We note that the peak wavelength positions of spectral features for these measurements do not always agree \citep{tamanai06}. This is due, for example, the effects of the inclusion of KBr in the measured sample or due to differences in the grain shape.

The circumstellar environment where we can observe debris from planetesimals are debris disks. These are dusty disks or rings around main-sequence stars, often with planets. These disks form from proto-planetary disks when the gas and small dust grains in the disk have been lost. These disks still show emission of small dust grains, but this dust needs to be continually replenished by collisions between planetesimals \citep{calvet05, wyatt07}. Interesting about these systems is that we know that the dust comes from planetesimals and thus traces the composition of these bodies. 

\cite{olofsson12} modelled several so called warm debris disks. \textcolor{\refrevisions1}{These systems have dust close enough to the star for it to be at least several hundreds of Kelvin (for example \cite{olofsson12} derive dust temperatures of 200-1500 K at the inner-edge of the disk, depending on the dust geometry, central temperature, dust composition, and other system parameters)}. This makes that these objects have infrared excesses on top of the stellar spectrum starting at $\sim$5-10 \mic \citep{bryden06, chen06}. These warm debris disks are rare since most debris disks are Kuiper-belt-type disks, having no to little emission from warm dust \citep{liseau10, lohne12}. For several of the warm debris disks \cite{olofsson12} finds emission features of olivine (and possible pyroxene). Modelling of the mid-IR spectra \cite{olofsson12} report that the olivine has an iron concentration of $\sim$20~\%. For their analysis they used the aerosol measurements of \cite{tamanai06}, which is a good comparison with the free floating grains in these disks. \cite{olofsson12} also found that the disks only contain minor amounts of pyroxene, where the Py/(Ol+Py) ratio ranged from 0 to 0.2. Thus for these warm debris disks it seems the olivine grains are rich in iron, but the pyroxene abundance is not high. Comparing these results to the meteorite groups discussed in this work, the olivine and pyroxene in these warm debris disk systems compares best to the carbonaceous chondrites type$\geq$3 or possibly pyroxene poor ordinary chondrites.

Proto-planetary disks around pre-main-sequence stars contain both dust and gas and these disks are the formation place of planets and planetesimals. In these systems the dust is probably not altered by parent-body processes, but some of the disks could be transitioning into a debris disk and in that case planetesimal debris could be present. \textcolor{\refrevisions1}{Furthermore, the spectra of chondrites presented in this work can help us understand the dust in proto-planetary disks since these dust grains are the building blocks of ordinary chondrites (matrix, chondrules and calcium aluminum inclusions).} Studies of a large set of mid-IR spectra of disks around T Tauri \citep{watson09} and Herbig Ae/Be \citep{juhasz10, maaskant14} stars showed that these disks can have a variety of olivine and pyroxene abundances, ranging from only olivine to an equal mixture of the two. It is thought that the olivine, the most stable of the two, condenses from the gas in the inner parts of the disk while the pyroxene is formed in colder regions further out in the disk \citep{gail04, abraham09, harker02}. \cite{vanboekel04} indeed showed, using interferometrical data, that for the proto-planetary disk HD~142527 the pyroxene over olivine ratio increases radially outwards. In contrast to this \cite{juhasz10} showed, based on studying mid-IR spectra, that for a large set of proto-planetary disks the enstatite is predominantly located in the inner disk, while the olivine is located in the outer parts of the disk. 

From the analysis of the mid-IR features it has been reported that the olivine is very magnesium rich in these proto-planetary disks and more recent studies including far-IR spectra concluded that the iron content of the olivine was less than 2\% \citep{sturm13, maaskant14}. These studies have been done using cross sections obtained from optical constants using model calculations. These cross section calculations simulate grains in a vacuum, but since the models require one to assume a certain shape, they might not be as good a comparison to astronomical dust as the aerosol measurements used in the study of \cite{olofsson12}. Besides the mid-IR astronomical spectra these studies also used the Herschel far-IR spectra with an olivine resonance at 69 \mic band. The low iron concentration of the olivine and the presence of significant amounts of pyroxene makes that the dust in these proto-planetary disks resembles the carbonaceous chondrites of type-2, but one has to keep in mind that the dust in these disks likely does not come from collisions, but from gas phase condensation and/or annealing.

\section{Conclusions}
In this work we showed that, and how, the different groups of meteorites can be distinguished when they would be spectroscopically studied in the form of debris. Since the different groups of meteorites can be traced back to parent bodies with different properties we can spectroscopically probe planetesimal properties (like size and its thermal and igneous evolution). The minerals of different meteorite groups can be spectroscopically distinguished by 1) the pyroxene-olivine ratio and 2) the iron in the olivine. These two properties can be measured in the mid-IR spectra by measuring feature strength ratios and peak shifts.

\bibliographystyle{elsarticle-harv}
\bibliography{references}

\begin{thebibliography}{65}
\expandafter\ifx\csname natexlab\endcsname\relax\def\natexlab#1{#1}\fi
\expandafter\ifx\csname url\endcsname\relax
  \def\url#1{\texttt{#1}}\fi
\expandafter\ifx\csname urlprefix\endcsname\relax\def\urlprefix{URL }\fi

\bibitem[{{{\'A}brah{\'a}m} et~al.(2009){{\'A}brah{\'a}m}, {Juh{\'a}sz},
  {Dullemond}, {K{\'o}sp{\'a}l}, {van Boekel}, {Bouwman}, {Henning},
  {Mo{\'o}r}, {Mosoni}, {Sicilia-Aguilar}, and {Sipos}}]{abraham09}
{{\'A}brah{\'a}m}, P., {Juh{\'a}sz}, A., {Dullemond}, C.~P., {K{\'o}sp{\'a}l},
  {\'A}., {van Boekel}, R., {Bouwman}, J., {Henning}, T., {Mo{\'o}r}, A.,
  {Mosoni}, L., {Sicilia-Aguilar}, A., {Sipos}, N., May 2009. {Episodic
  formation of cometary material in the outburst of a young Sun-like star}.
  \nat 459, 224--226.

\bibitem[{{Beck} et~al.(2014){Beck}, {Garenne}, {Quirico}, {Bonal},
  {Montes-Hernandez}, {Moynier}, and {Schmitt}}]{beck14}
{Beck}, P., {Garenne}, A., {Quirico}, E., {Bonal}, L., {Montes-Hernandez}, G.,
  {Moynier}, F., {Schmitt}, B., Feb. 2014. {Transmission infrared spectra (2-25
  {$\mu$}m) of carbonaceous chondrites (CI, CM, CV-CK, CR, C2 ungrouped):
  Mineralogy, water, and asteroidal processes}. \icarus 229, 263--277.

\bibitem[{{Bischoff} et~al.(1993){Bischoff}, {Palme}, {Schultz}, {Weber},
  {Weber}, and {Spettel}}]{bischoff93}
{Bischoff}, A., {Palme}, H., {Schultz}, L., {Weber}, D., {Weber}, H.~W.,
  {Spettel}, B., Jun. 1993. {ACFER 182 and paired samples, an iron-rich
  carbonaceous chondrite - Similarities with ALH85085 and relationship to CR
  chondrites}. \gca 57, 2631--2648.

\bibitem[{{Bowey} et~al.(2007){Bowey}, {Morlok}, {K{\"o}hler}, and
  {Grady}}]{bowey07}
{Bowey}, J.~E., {Morlok}, A., {K{\"o}hler}, M., {Grady}, M., Apr. 2007.
  {2-16{$\mu$}m spectroscopy of micron-sized enstatite
  (Mg,Fe)$_{2}$Si$_{2}$O$_{6}$ silicates from primitive chondritic meteorites}.
  \mnras 376, 1367--1374.

\bibitem[{{Bradley} et~al.(1999){Bradley}, {Keller}, {Snow}, {Hanner}, {Flynn},
  {Gezo}, {Clemett}, {Brownlee}, and {Bowey}}]{bradley99}
{Bradley}, J.~P., {Keller}, L.~P., {Snow}, T.~P., {Hanner}, M.~S., {Flynn},
  G.~J., {Gezo}, J.~C., {Clemett}, S.~J., {Brownlee}, D.~E., {Bowey}, J.~E.,
  Sep. 1999. {An infrared spectral match between GEMS and interstellar grains}.
  Science 285.

\bibitem[{{Brearley} and {Jones}(1998)}]{brearleyjones98}
{Brearley}, A., {Jones}, R., 1998. {Chondritic meteorites}. Planetary Materials
  398, 3--1--3.

\bibitem[{{Bryden} et~al.(2006){Bryden}, {Beichman}, {Trilling}, {Rieke},
  {Holmes}, {Lawler}, {Stapelfeldt}, {Werner}, {Gautier}, {Blaylock}, {Gordon},
  {Stansberry}, and {Su}}]{bryden06}
{Bryden}, G., {Beichman}, C.~A., {Trilling}, D.~E., {Rieke}, G.~H., {Holmes},
  E.~K., {Lawler}, S.~M., {Stapelfeldt}, K.~R., {Werner}, M.~W., {Gautier},
  T.~N., {Blaylock}, M., {Gordon}, K.~D., {Stansberry}, J.~A., {Su}, K.~Y.~L.,
  Jan. 2006. {Frequency of Debris Disks around Solar-Type Stars: First Results
  from a Spitzer MIPS Survey}. \apj 636, 1098--1113.

\bibitem[{{Calvet} et~al.(2005){Calvet}, {D'Alessio}, {Watson},
  {Franco-Hern{\'a}ndez}, {Furlan}, {Green}, {Sutter}, {Forrest}, {Hartmann},
  {Uchida}, {Keller}, {Sargent}, {Najita}, {Herter}, {Barry}, and
  {Hall}}]{calvet05}
{Calvet}, N., {D'Alessio}, P., {Watson}, D.~M., {Franco-Hern{\'a}ndez}, R.,
  {Furlan}, E., {Green}, J., {Sutter}, P.~M., {Forrest}, W.~J., {Hartmann}, L.,
  {Uchida}, K.~I., {Keller}, L.~D., {Sargent}, B., {Najita}, J., {Herter},
  T.~L., {Barry}, D.~J., {Hall}, P., Sep. 2005. {Disks in Transition in the
  Taurus Population: Spitzer IRS Spectra of GM Aurigae and DM Tauri}. \apjl
  630, L185--L188.

\bibitem[{{Chen} et~al.(2007){Chen}, {Li}, {Bohac}, {Kim}, {Watson}, {van
  Cleve}, {Houck}, {Stapelfeldt}, {Werner}, {Rieke}, {Su}, {Marengo},
  {Backman}, {Beichman}, and {Fazio}}]{chen07}
{Chen}, C.~H., {Li}, A., {Bohac}, C., {Kim}, K.~H., {Watson}, D.~M., {van
  Cleve}, J., {Houck}, J., {Stapelfeldt}, K., {Werner}, M.~W., {Rieke}, G.,
  {Su}, K., {Marengo}, M., {Backman}, D., {Beichman}, C., {Fazio}, G., Sep.
  2007. {The Dust and Gas Around {$\beta$} Pictoris}. \apj 666, 466--474.

\bibitem[{{Chen} et~al.(2006){Chen}, {Sargent}, {Bohac}, {Kim},
  {Leibensperger}, {Jura}, {Najita}, {Forrest}, {Watson}, {Sloan}, and
  {Keller}}]{chen06}
{Chen}, C.~H., {Sargent}, B.~A., {Bohac}, C., {Kim}, K.~H., {Leibensperger},
  E., {Jura}, M., {Najita}, J., {Forrest}, W.~J., {Watson}, D.~M., {Sloan},
  G.~C., {Keller}, L.~D., Sep. 2006. {Spitzer IRS Spectroscopy of
  IRAS-discovered Debris Disks}. \apjs 166, 351--377.

\bibitem[{{Chihara} et~al.(2002){Chihara}, {Koike}, {Tsuchiyama}, {Tachibana},
  and {Sakamoto}}]{chihara02}
{Chihara}, H., {Koike}, C., {Tsuchiyama}, A., {Tachibana}, S., {Sakamoto}, D.,
  Aug. 2002. {Compositional dependence of infrared absorption spectra of
  crystalline silicates. I. Mg-Fe pyroxenes}. \aap 391, 267--273.

\bibitem[{{Consolmagno} and {Drake}(1977)}]{consolmagno77}
{Consolmagno}, G.~J., {Drake}, M.~J., Sep. 1977. {Composition and evolution of
  the eucrite parent body - Evidence from rare earth elements}. \gca 41,
  1271--1282.

\bibitem[{{de Vries} et~al.(2012){de Vries}, {Acke}, {Blommaert}, {Waelkens},
  {Waters}, {Vandenbussche}, {Min}, {Olofsson}, {Dominik}, {Decin}, {Barlow},
  {Brandeker}, {di Francesco}, {Glauser}, {Greaves}, {Harvey}, {Holland},
  {Ivison}, {Liseau}, {Pantin}, {Pilbratt}, {Royer}, and
  {Sibthorpe}}]{devries12}
{de Vries}, B.~L., {Acke}, B., {Blommaert}, J.~A.~D.~L., {Waelkens}, C.,
  {Waters}, L.~B.~F.~M., {Vandenbussche}, B., {Min}, M., {Olofsson}, G.,
  {Dominik}, C., {Decin}, L., {Barlow}, M.~J., {Brandeker}, A., {di Francesco},
  J., {Glauser}, A.~M., {Greaves}, J., {Harvey}, P.~M., {Holland}, W.~S.,
  {Ivison}, R.~J., {Liseau}, R., {Pantin}, E.~E., {Pilbratt}, G.~L., {Royer},
  P., {Sibthorpe}, B., Oct. 2012. {Comet-like mineralogy of olivine crystals in
  an extrasolar proto-Kuiper belt}. \nat 490, 74--76.

\bibitem[{{de Vries} et~al.(2015){de Vries}, {Maaskant}, {Min}, {Lombaert},
  {Waters}, and {Blommaert}}]{devries15}
{de Vries}, B.~L., {Maaskant}, K.~M., {Min}, M., {Lombaert}, R., {Waters},
  L.~B.~F.~M., {Blommaert}, J.~A.~D.~L., Apr. 2015. {Micron-sized forsterite
  grains in the pre-planetary nebula of IRAS 17150-3224. Searching for clues to
  the mysterious evolution of massive AGB stars}. \aap 576, A98.

\bibitem[{{Fowler} et~al.(1994){Fowler}, {Papike}, {Spilde}, and
  {Shearer}}]{fowler94}
{Fowler}, G.~W., {Papike}, J.~J., {Spilde}, M.~N., {Shearer}, C.~K., Sep. 1994.
  {Diogenites as asteroidal cumulates: Insights from orthopyroxene major and
  minor element chemistry}. \gca 58, 3921--3929.

\bibitem[{{Gail}(2004)}]{gail04}
{Gail}, H.-P., Jan. 2004. {Radial mixing in protoplanetary accretion disks. IV.
  Metamorphosis of the silicate dust complex}. \aap 413, 571--591.

\bibitem[{{Gail} and {Sedlmayr}(1999)}]{gailsedl99}
{Gail}, H.-P., {Sedlmayr}, E., Jul. 1999. {Mineral formation in stellar winds.
  I. Condensation sequence of silicate and iron grains in stationary oxygen
  rich outflows}. \aap 347, 594--616.

\bibitem[{{Harker} and {Desch}(2002)}]{harker02}
{Harker}, D.~E., {Desch}, S.~J., Feb. 2002. {Annealing of Silicate Dust by
  Nebular Shocks at 10 AU}. \apjl 565, L109--L112.

\bibitem[{{Hutchison}(2004)}]{hutchison04_book}
{Hutchison}, R., Oct. 2004. {Meteorites}. Cambridge University Press.

\bibitem[{{J{\" a}ger} et~al.(1998){J{\" a}ger}, {Molster}, {Dorschner},
  {Henning}, {Mutschke}, and {Waters}}]{jager98}
{J{\" a}ger}, C., {Molster}, F.~J., {Dorschner}, J., {Henning}, T., {Mutschke},
  H., {Waters}, L.~B.~F.~M., Nov. 1998. {Steps toward interstellar silicate
  mineralogy. IV. The crystalline revolution}. \aap 339, 904--916.

\bibitem[{{Juh{\'a}sz} et~al.(2010){Juh{\'a}sz}, {Bouwman}, {Henning}, {Acke},
  {van den Ancker}, {Meeus}, {Dominik}, {Min}, {Tielens}, and
  {Waters}}]{juhasz10}
{Juh{\'a}sz}, A., {Bouwman}, J., {Henning}, T., {Acke}, B., {van den Ancker},
  M.~E., {Meeus}, G., {Dominik}, C., {Min}, M., {Tielens}, A.~G.~G.~M.,
  {Waters}, L.~B.~F.~M., Sep. 2010. {Dust Evolution in Protoplanetary Disks
  Around Herbig Ae/Be Stars - the Spitzer View}. \apj 721, 431--455.

\bibitem[{{Kallemeyn} et~al.(1994){Kallemeyn}, {Rubin}, and
  {Wasson}}]{kallemeyn94}
{Kallemeyn}, G.~W., {Rubin}, A.~E., {Wasson}, J.~T., Jul. 1994. {The
  compositional classification of chondrites: VI. The CR carbonaceous chondrite
  group}. \gca 58, 2873--2888.

\bibitem[{{Kemper} et~al.(2004){Kemper}, {Vriend}, and {Tielens}}]{kemper04}
{Kemper}, F., {Vriend}, W.~J., {Tielens}, A.~G.~G.~M., Jul. 2004. {The Absence
  of Crystalline Silicates in the Diffuse Interstellar Medium}. \apj 609,
  826--837.

\bibitem[{{Kessel} et~al.(2007){Kessel}, {Beckett}, and {Stolper}}]{kessel07}
{Kessel}, R., {Beckett}, J.~R., {Stolper}, E.~M., Apr. 2007. {The thermal
  history of equilibrated ordinary chondrites and the relationship between
  textural maturity and temperature}. \gca 71, 1855--1881.

\bibitem[{{Kessler-Silacci} et~al.(2006){Kessler-Silacci}, {Augereau},
  {Dullemond}, {Geers}, {Lahuis}, {Evans}, {van Dishoeck}, {Blake}, {Brown},
  {J{\o}rgensen}, {Knez}, and {Pontoppidan}}]{spitzer1}
{Kessler-Silacci}, J., {Augereau}, J.-C., {Dullemond}, C.~P., {Geers}, V.,
  {Lahuis}, F., {Evans}, II, N.~J., {van Dishoeck}, E.~F., {Blake},
  G.~A.~and{Boogert}, A.~C.~A., {Brown}, J., {J{\o}rgensen}, J.~K., {Knez}, C.,
  {Pontoppidan}, K.~M., Mar. 2006. {c2d Spitzer IRS Spectra of Disks around T
  Tauri Stars. I. Silicate Emission and Grain Growth}. \apj 639, 275--291.

\bibitem[{{Koike} et~al.(2003){Koike}, {Chihara}, {Tsuchiyama}, {Suto},
  {Sogawa}, and {Okuda}}]{koike03}
{Koike}, C., {Chihara}, H., {Tsuchiyama}, A., {Suto}, H., {Sogawa}, H.,
  {Okuda}, H., Mar. 2003. {Compositional dependence of infrared absorption
  spectra of crystalline silicate. II. Natural and synthetic olivines}. \aap
  399, 1101--1107.

\bibitem[{{Liseau} et~al.(2010){Liseau}, {Eiroa}, {Fedele}, {Augereau},
  {Olofsson}, {Gonz{\'a}lez}, {Maldonado}, {Montesinos}, {Mora}, {Absil},
  {Ardila}, {Barrado}, {Bayo}, {Beichman}, {Bryden}, {Danchi}, {Del Burgo},
  {Ertel}, {Fridlund}, {Heras}, {Krivov}, {Launhardt}, {Lebreton}, {L{\"o}hne},
  {Marshall}, {Meeus}, {M{\"u}ller}, {Pilbratt}, {Roberge}, {Rodmann},
  {Solano}, {Stapelfeldt}, {Th{\'e}bault}, {White}, and {Wolf}}]{liseau10}
{Liseau}, R., {Eiroa}, C., {Fedele}, D., {Augereau}, J.-C., {Olofsson}, G.,
  {Gonz{\'a}lez}, B., {Maldonado}, J., {Montesinos}, B., {Mora}, A., {Absil},
  O., {Ardila}, D., {Barrado}, D., {Bayo}, A., {Beichman}, C.~A., {Bryden}, G.,
  {Danchi}, W.~C., {Del Burgo}, C., {Ertel}, S., {Fridlund}, C.~W.~M., {Heras},
  A.~M., {Krivov}, A.~V., {Launhardt}, R., {Lebreton}, J., {L{\"o}hne}, T.,
  {Marshall}, J.~P., {Meeus}, G., {M{\"u}ller}, S., {Pilbratt}, G.~L.,
  {Roberge}, A., {Rodmann}, J., {Solano}, E., {Stapelfeldt}, K.~R.,
  {Th{\'e}bault}, P., {White}, G.~J., {Wolf}, S., Jul. 2010. {Resolving the
  cold debris disc around a planet-hosting star . PACS photometric imaging
  observations of q$^{1}$ Eridani (HD 10647, HR 506)}. \aap 518, L132.

\bibitem[{{L{\"o}hne} et~al.(2012){L{\"o}hne}, {Augereau}, {Ertel}, {Marshall},
  {Eiroa}, {Mora}, {Absil}, {Stapelfeldt}, {Th{\'e}bault}, {Bayo}, {Del Burgo},
  {Danchi}, {Krivov}, {Lebreton}, {Letawe}, {Magain}, {Maldonado},
  {Montesinos}, {Pilbratt}, {White}, and {Wolf}}]{lohne12}
{L{\"o}hne}, T., {Augereau}, J.-C., {Ertel}, S., {Marshall}, J.~P., {Eiroa},
  C., {Mora}, A., {Absil}, O., {Stapelfeldt}, K., {Th{\'e}bault}, P., {Bayo},
  A., {Del Burgo}, C., {Danchi}, W., {Krivov}, A.~V., {Lebreton}, J., {Letawe},
  G., {Magain}, P., {Maldonado}, J., {Montesinos}, B., {Pilbratt}, G.~L.,
  {White}, G.~J., {Wolf}, S., Jan. 2012. {Modelling the huge, Herschel-resolved
  debris ring around HD 207129}. \aap 537, A110.

\bibitem[{{Maaskant} et~al.(2015){Maaskant}, {de Vries}, {Min}, {Waters},
  {Dominik}, {Molster}, and {Tielens}}]{maaskant14}
{Maaskant}, K.~M., {de Vries}, B.~L., {Min}, M., {Waters}, L.~B.~F.~M.,
  {Dominik}, C., {Molster}, F., {Tielens}, A.~G.~G.~M., Feb. 2015. {Location
  and sizes of forsterite grains in protoplanetary disks. Interpretation from
  the Herschel DIGIT programme}. \aap 574, A140.

\bibitem[{{Markwick-Kemper} et~al.(2007){Markwick-Kemper}, {Gallagher},
  {Hines}, and {Bouwman}}]{kemper07}
{Markwick-Kemper}, F., {Gallagher}, S.~C., {Hines}, D.~C., {Bouwman}, J., Oct.
  2007. {Dust in the Wind: Crystalline Silicates, Corundum, and Periclase in PG
  2112+059}. \apjl 668, L107--L110.

\bibitem[{{McCord} et~al.(1970){McCord}, {Adams}, and {Johnson}}]{mccord70}
{McCord}, T.~B., {Adams}, J.~B., {Johnson}, T.~V., Jun. 1970. {Asteroid Vesta:
  Spectral Reflectivity and Compositional Implications}. Science 168,
  1445--1447.

\bibitem[{{Meeus} et~al.(2001){Meeus}, {Waters}, {Bouwman}, {van den Ancker},
  {Waelkens}, and {Malfait}}]{meeus01}
{Meeus}, G., {Waters}, L.~B.~F.~M., {Bouwman}, J., {van den Ancker}, M.~E.,
  {Waelkens}, C., {Malfait}, K., Jan. 2001. {ISO spectroscopy of circumstellar
  dust in 14 Herbig Ae/Be systems: Towards an understanding of dust
  processing}. \aap 365, 476--490.

\bibitem[{{Min} et~al.(2008){Min}, {Hovenier}, {Waters}, and {de
  Koter}}]{min08}
{Min}, M., {Hovenier}, J.~W., {Waters}, L.~B.~F.~M., {de Koter}, A., Oct. 2008.
  {The infrared emission spectra of compositionally inhomogeneous aggregates
  composed of irregularly shaped constituents}. \aap 489, 135--141.

\bibitem[{{Min} et~al.(2007){Min}, {Waters}, {de Koter}, {Hovenier}, {Keller},
  and {Markwick-Kemper}}]{min07_ISMgrains}
{Min}, M., {Waters}, L.~B.~F.~M., {de Koter}, A., {Hovenier}, J.~W., {Keller},
  L.~P., {Markwick-Kemper}, F., Feb. 2007. {The shape and composition of
  interstellar silicate grains}. \aap 462, 667--676.

\bibitem[{{Molster} et~al.(2003){Molster}, {Demyk}, {D'Hendecourt}, and
  {Bradley}}]{molster03}
{Molster}, F.~J., {Demyk}, A., {D'Hendecourt}, L., {Bradley}, J.~P., Mar. 2003.
  {The First 2-50 $\mu$m Infrared Spectrum of an Interplanetary Dust Particle
  (IDP)}. In: {Mackwell}, S., {Stansbery}, E. (Eds.), Lunar and Planetary
  Science Conference. Vol.~34 of Lunar and Planetary Inst. Technical Report. p.
  1148.

\bibitem[{{Molster} et~al.(2002){Molster}, {Waters}, {Tielens}, and
  {Barlow}}]{mol02_1}
{Molster}, F.~J., {Waters}, L.~B.~F.~M., {Tielens}, A.~G.~G.~M., {Barlow},
  M.~J., Jan. 2002. {Crystalline silicate dust around evolved stars. I. The
  sample stars}. \aap 382, 184--221.

\bibitem[{{Morlok} et~al.(2012){Morlok}, {Koike}, {Tomeoka}, {Mason}, {Lisse},
  {Anand}, and {Grady}}]{morlok12}
{Morlok}, A., {Koike}, C., {Tomeoka}, K., {Mason}, A., {Lisse}, C., {Anand},
  M., {Grady}, M., May 2012. {Mid-infrared spectra of differentiated meteorites
  (achondrites): Comparison with astronomical observations of dust in
  protoplanetary and debris disks}. \icarus 219, 48--56.

\bibitem[{{Morlok} et~al.(2010){Morlok}, {Koike}, {Tomioka}, {Mann}, and
  {Tomeoka}}]{morlok10}
{Morlok}, A., {Koike}, C., {Tomioka}, N., {Mann}, I., {Tomeoka}, K., May 2010.
  {Mid-infrared spectra of the shocked Murchison CM chondrite: Comparison with
  astronomical observations of dust in debris disks}. \icarus 207, 45--53.

\bibitem[{{Morlok} et~al.(2014{\natexlab{a}}){Morlok}, {Lisse}, {Mason},
  {Bullock}, and {Grady}}]{morlok14a}
{Morlok}, A., {Lisse}, C.~M., {Mason}, A.~B., {Bullock}, E.~S., {Grady}, M.~M.,
  Mar. 2014{\natexlab{a}}. {Mid-infrared spectroscopy of components in
  chondrites: Search for processed materials in young Solar Systems and
  comets}. \icarus 231, 338--355.

\bibitem[{{Morlok} et~al.(2014{\natexlab{b}}){Morlok}, {Mason}, {Anand},
  {Lisse}, {Bullock}, and {Grady}}]{morlok14b}
{Morlok}, A., {Mason}, A.~B., {Anand}, M., {Lisse}, C.~M., {Bullock}, E.~S.,
  {Grady}, M.~M., Sep. 2014{\natexlab{b}}. {Dust from collisions: A way to
  probe the composition of exo-planets?} \icarus 239, 1--14.

\bibitem[{{Nakamura} et~al.(2011){Nakamura}, {Noguchi}, {Tanaka}, {Zolensky},
  {Kimura}, {Tsuchiyama}, {Nakato}, {Ogami}, {Ishida}, {Uesugi}, {Yada},
  {Shirai}, {Fujimura}, {Okazaki}, {Sandford}, {Ishibashi}, {Abe}, {Okada},
  {Ueno}, {Mukai}, {Yoshikawa}, and {Kawaguchi}}]{nakamura11}
{Nakamura}, T., {Noguchi}, T., {Tanaka}, M., {Zolensky}, M.~E., {Kimura}, M.,
  {Tsuchiyama}, A., {Nakato}, A., {Ogami}, T., {Ishida}, H., {Uesugi}, M.,
  {Yada}, T., {Shirai}, K., {Fujimura}, A., {Okazaki}, R., {Sandford}, S.~A.,
  {Ishibashi}, Y., {Abe}, M., {Okada}, T., {Ueno}, M., {Mukai}, T.,
  {Yoshikawa}, M., {Kawaguchi}, J., Aug. 2011. {Itokawa Dust Particles: A
  Direct Link Between S-Type Asteroids and Ordinary Chondrites}. Science 333,
  1113--.

\bibitem[{{Olofsson} et~al.(2012){Olofsson}, {Juh{\'a}sz}, {Henning},
  {Mutschke}, {Tamanai}, {Mo{\'o}r}, and {{\'A}brah{\'a}m}}]{olofsson12}
{Olofsson}, J., {Juh{\'a}sz}, A., {Henning}, T., {Mutschke}, H., {Tamanai}, A.,
  {Mo{\'o}r}, A., {{\'A}brah{\'a}m}, P., Jun. 2012. {Transient dust in warm
  debris disks. Detection of Fe-rich olivine grains}. \aap 542, A90.

\bibitem[{{Ordal} et~al.(1988){Ordal}, {Bell}, {Alexander}, {Newquist}, and
  {Querry}}]{ordal88}
{Ordal}, M.~A., {Bell}, R.~J., {Alexander}, Jr., R.~W., {Newquist}, L.~A.,
  {Querry}, M.~R., Mar. 1988. {Optical properties of Al, Fe, Ti, Ta, W, and Mo
  at submillimeter wavelengths}. \ao 27, 1203--1209.

\bibitem[{{Rubin} et~al.(1988){Rubin}, {Wang}, {Kallemeyn}, and
  {Wasson}}]{rubin88}
{Rubin}, A.~E., {Wang}, D., {Kallemeyn}, G.~W., {Wasson}, J.~T., Mar. 1988.
  {The Ningqiang meteorite - Classification and petrology of an anomalous CV
  chondrite}. Meteoritics 23, 13--23.

\bibitem[{{Russell} et~al.(2015){Russell}, {McSween}, {Jaumann}, and
  {Raymond}}]{russell15_dawn}
{Russell}, C.~T., {McSween}, H.~Y., {Jaumann}, R., {Raymond}, C.~A., 2015. {The
  Dawn Mission to Vesta and Ceres}. pp. 419--432.

\bibitem[{{Salisbury} et~al.(1991){Salisbury}, {Walter}, {Vergo}, and
  {D'Aria}}]{salisbury91}
{Salisbury}, J.~W., {Walter}, L.~S., {Vergo}, N., {D'Aria}, D., 1991. {Infrared
  (2.1-25 micron) Spectra of Minerals}. Johns Hopkins University Press.

\bibitem[{{Sandford} and {Walker}(1985)}]{sandford85}
{Sandford}, S.~A., {Walker}, R.~M., Apr. 1985. {Laboratory infrared
  transmission spectra of individual interplanetary dust particles from 2.5 to
  25 microns}. \apj 291, 838--851.

\bibitem[{{Sargent} et~al.(2009){Sargent}, {Forrest}, {Tayrien}, {McClure},
  {Watson}, {Sloan}, {Li}, {Manoj}, {Bohac}, {Furlan}, {Kim}, and
  {Green}}]{sargent09}
{Sargent}, B.~A., {Forrest}, W.~J., {Tayrien}, C., {McClure}, M.~K., {Watson},
  D.~M., {Sloan}, G.~C., {Li}, A., {Manoj}, P., {Bohac}, C.~J., {Furlan}, E.,
  {Kim}, K.~H., {Green}, J.~D., Jun. 2009. {Dust Processing and Grain Growth in
  Protoplanetary Disks in the Taurus-Auriga Star-Forming Region}. \apjs 182,
  477--508.

\bibitem[{{Servoin} and {Piriou}(1973)}]{servoin73}
{Servoin}, J.~L., {Piriou}, B., 1973. {Infrared reflectivity and Raman
  scattering of Mg2SiO4 single crystal}. \rm Phys. Status Solidi (B) 55,
  677--686.

\bibitem[{{Sogawa} and {Kozasa}(1999)}]{sogawa99}
{Sogawa}, H., {Kozasa}, T., May 1999. {On the Origin of Crystalline Silicate in
  Circumstellar Envelopesof Oxygen-rich Asymptotic Giant Branch Stars}. \apjl
  516, L33--L36.

\bibitem[{{Spoon} et~al.(2006){Spoon}, {Tielens}, {Armus}, {Sloan}, {Sargent},
  {Cami}, {Charmandaris}, {Houck}, and {Soifer}}]{spoon06}
{Spoon}, H.~W.~W., {Tielens}, A.~G.~G.~M., {Armus}, L., {Sloan}, G.~C.,
  {Sargent}, B., {Cami}, J., {Charmandaris}, V., {Houck}, J.~R., {Soifer},
  B.~T., Feb. 2006. {The Detection of Crystalline Silicates in Ultraluminous
  Infrared Galaxies}. \apj 638, 759--765.

\bibitem[{{Sturm} et~al.(2013){Sturm}, {Bouwman}, {Henning}, {Evans}, {Waters},
  {van Dishoeck}, {Green}, {Olofsson}, {Meeus}, {Maaskant}, {Dominik},
  {Augereau}, {Mulders}, {Acke}, {Merin}, and {Herczeg}}]{sturm13}
{Sturm}, B., {Bouwman}, J., {Henning}, T., {Evans}, N.~J., {Waters},
  L.~B.~F.~M., {van Dishoeck}, E.~F., {Green}, J.~D., {Olofsson}, J., {Meeus},
  G., {Maaskant}, K., {Dominik}, C., {Augereau}, J.~C., {Mulders}, G.~D.,
  {Acke}, B., {Merin}, B., {Herczeg}, G.~J., May 2013. {The 69 {$\mu$}m
  forsterite band in spectra of protoplanetary disks. Results from the Herschel
  DIGIT programme}. \aap 553, A5.

\bibitem[{{Sylvester} et~al.(1999){Sylvester}, {Kemper}, {Barlow}, {de Jong},
  {Waters}, {Tielens}, and {Omont}}]{syl99}
{Sylvester}, R.~J., {Kemper}, F., {Barlow}, M.~J., {de Jong}, T., {Waters},
  L.~B.~F.~M., {Tielens}, A.~G.~G.~M., {Omont}, A., Dec. 1999. {2.4-197 mu m
  spectroscopy of OH/IR stars: the IR characteristics of circumstellar dust in
  O-rich environments}. \aap 352, 587--599.

\bibitem[{{Takeda}(1997)}]{takeda97}
{Takeda}, H., Nov. 1997. {Mineralogical records of early planetary processes on
  the HED parent body with reference to Vesta}. Meteoritics and Planetary
  Science 32.

\bibitem[{{Tamanai} et~al.(2009){Tamanai}, {Mutschke}, and {Blum}}]{tamanai06}
{Tamanai}, A., {Mutschke}, H., {Blum}, J., Dec. 2009. {IR Spectroscopic
  Measurements of Free-Flying Silicate Dust Grains}. In: {Henning}, T.,
  {Gr{\"u}n}, E., {Steinacker}, J. (Eds.), Cosmic Dust - Near and Far. Vol. 414
  of Astronomical Society of the Pacific Conference Series. p. 438.

\bibitem[{{Tielens} et~al.(1998){Tielens}, {Waters}, {Molster}, and
  {Justtanont}}]{tielens98}
{Tielens}, A.~G.~G.~M., {Waters}, L.~B.~F.~M., {Molster}, F.~J., {Justtanont},
  K., 1998. {Circumstellar Silicate Mineralogy}. \apss 255, 415--426.

\bibitem[{{van Boekel} et~al.(2004){van Boekel}, {Min}, {Leinert}, {Waters},
  {Richichi}, {Chesneau}, {Dominik}, {Jaffe}, {Dutrey}, {Graser}, {Henning},
  {de Jong}, {K{\"o}hler}, {de Koter}, {Lopez}, {Malbet}, {Morel}, {Paresce},
  {Perrin}, {Preibisch}, {Przygodda}, {Sch{\"o}ller}, and
  {Wittkowski}}]{vanboekel04}
{van Boekel}, R., {Min}, M., {Leinert}, C., {Waters}, L.~B.~F.~M., {Richichi},
  A., {Chesneau}, O., {Dominik}, C., {Jaffe}, W., {Dutrey}, A., {Graser}, U.,
  {Henning}, T., {de Jong}, J., {K{\"o}hler}, R., {de Koter}, A., {Lopez}, B.,
  {Malbet}, F., {Morel}, S., {Paresce}, F., {Perrin}, G., {Preibisch}, T.,
  {Przygodda}, F., {Sch{\"o}ller}, M., {Wittkowski}, M., Nov. 2004. {The
  building blocks of planets within the `terrestrial' region of protoplanetary
  disks}. \nat 432, 479--482.

\bibitem[{{Van Schmus}(1969)}]{vanschmus69}
{Van Schmus}, W., 1969. {Mineralogy, petrology and classification of types 3
  and 4 carbonaceous chondrites.} Meteorite Research 480, 480--91.

\bibitem[{{Waelkens} et~al.(1996){Waelkens}, {Waters}, {de Graauw}, {Huygen},
  {Malfait}, {Plets}, {Vandenbussche}, {Beintema}, {Habing}, {Heras}, {Kester},
  {Lahuis}, {Morris}, {Roelfsema}, {Salama}, {Siebenmorgen}, {Trams}, {van der
  Bliek}, {Valentijn}, and {Wesselius}}]{waelkens96}
{Waelkens}, C., {Waters}, L.~B.~F.~M., {de Graauw}, M.~S., {Huygen}, E.,
  {Malfait}, K., {Plets}, H., {Vandenbussche}, B., {Beintema},
  D.~A.~and{Boxhoorn}, D.~R., {Habing}, H.~J., {Heras}, A.~M., {Kester},
  D.~J.~M., {Lahuis}, F., {Morris}, P.~W., {Roelfsema}, P.~R., {Salama}, A.,
  {Siebenmorgen}, R., {Trams}, N.~R., {van der Bliek}, N.~R., {Valentijn},
  E.~A., {Wesselius}, P.~R., Nov. 1996. {SWS observations of young
  main-sequence stars with dusty circumstellar disks.} \aap 315, L245--L248.

\bibitem[{{Waters} et~al.(1996){Waters}, {Molster}, {de Jong}, {Beintema},
  {Waelkens}, {Boogert}, {Boxhoorn}, {de Graauw}, {Feuchtgruber}, {Genzel},
  {Helmich}, {Heras}, {Huygen}, {Izumiura}, {Justtanont}, {Kester}, {Lahuis},
  {Lamers}, {Leech}, {Loup}, {Lutz}, {Morris}, {Price}, {Roelfsema}, {Salama},
  {Schaeidt}, {Tielens}, {Trams}, {Valentijn}, {Vandenbussche}, {van den
  Ancker}, {van Dishoeck}, {Van Winckel}, {Wesselius}, and {Young}}]{waters96}
{Waters}, L.~B.~F.~M., {Molster}, F.~J., {de Jong}, T., {Beintema}, D.~A.,
  {Waelkens}, C., {Boogert}, A.~C.~A., {Boxhoorn}, D.~R., {de Graauw},
  T.~and{Drapatz}, S., {Feuchtgruber}, H., {Genzel}, R., {Helmich}, F.~P.,
  {Heras}, A.~M., {Huygen}, R., {Izumiura}, H., {Justtanont}, K., {Kester},
  D.~J.~M.~and{Kunze}, D., {Lahuis}, F., {Lamers}, H.~J.~G.~L.~M., {Leech},
  K.~J., {Loup}, C., {Lutz}, D., {Morris}, P.~W., {Price}, S.~D., {Roelfsema},
  P.~R., {Salama}, A., {Schaeidt}, S.~G., {Tielens}, A.~G.~G.~M., {Trams},
  N.~R., {Valentijn}, E.~A., {Vandenbussche}, B., {van den Ancker}, M.~E., {van
  Dishoeck}, E.~F., {Van Winckel}, H., {Wesselius}, P.~R., {Young}, E.~T., Nov.
  1996. {Mineralogy of oxygen-rich dust shells.} \aap 315, L361--L364.

\bibitem[{{Watson} et~al.(2009){Watson}, {Leisenring}, {Furlan}, {Bohac},
  {Sargent}, {Forrest}, {Calvet}, {Hartmann}, {Nordhaus}, {Green}, {Kim},
  {Sloan}, {Chen}, {Keller}, {d'Alessio}, {Najita}, {Uchida}, and
  {Houck}}]{watson09}
{Watson}, D.~M., {Leisenring}, J.~M., {Furlan}, E., {Bohac}, C.~J., {Sargent},
  B., {Forrest}, W.~J., {Calvet}, N., {Hartmann}, L., {Nordhaus}, J.~T.,
  {Green}, J.~D., {Kim}, K.~H., {Sloan}, G.~C., {Chen}, C.~H., {Keller}, L.~D.,
  {d'Alessio}, P., {Najita}, J., {Uchida}, K.~I., {Houck}, J.~R., Jan. 2009.
  {Crystalline Silicates and Dust Processing in the Protoplanetary Disks of the
  Taurus Young Cluster}. \apjs 180, 84--101.

\bibitem[{{Wooden}(2002)}]{wooden02}
{Wooden}, D.~H., Oct. 2002. {Comet Grains: Their IR Emission and Their Relation
  to ISm Grains}. Earth Moon and Planets 89, 247--287.

\bibitem[{{Wyatt}(2008)}]{wyatt08}
{Wyatt}, M.~C., Sep. 2008. {Evolution of Debris Disks}. \araa 46, 339--383.

\bibitem[{{Wyatt} et~al.(2007){Wyatt}, {Smith}, {Su}, {Rieke}, {Greaves},
  {Beichman}, and {Bryden}}]{wyatt07}
{Wyatt}, M.~C., {Smith}, R., {Su}, K.~Y.~L., {Rieke}, G.~H., {Greaves}, J.~S.,
  {Beichman}, C.~A., {Bryden}, G., Jul. 2007. {Steady State Evolution of Debris
  Disks around A Stars}. \apj 663, 365--382.

\bibitem[{{Zolensky} et~al.(2008){Zolensky}, {Nakamura-Messenger},
  {Rietmeijer}, {Leroux}, {Mikouchi}, {Ohsumi}, {Simon}, {Grossman}, {Stephan},
  {Weisberg}, {Velbel}, {Zega}, {Stroud}, {Tomeoka}, {Ohnishi}, {Tomioka},
  {Nakamura}, {Matrajt}, {Joswiak}, {Brownlee}, {Langenhorst}, {Krot},
  {Kearsley}, {Ishii}, {Graham}, {Dai}, {Chi}, {Bradley}, {Hagiya}, {Gounelle},
  and {Bridges}}]{zolensky08}
{Zolensky}, M., {Nakamura-Messenger}, K., {Rietmeijer}, F., {Leroux}, H.,
  {Mikouchi}, T., {Ohsumi}, K., {Simon}, S., {Grossman}, L., {Stephan}, T.,
  {Weisberg}, M., {Velbel}, M., {Zega}, T., {Stroud}, R., {Tomeoka}, K.,
  {Ohnishi}, I., {Tomioka}, N., {Nakamura}, T., {Matrajt}, G., {Joswiak}, D.,
  {Brownlee}, D., {Langenhorst}, F., {Krot}, A., {Kearsley}, A., {Ishii}, H.,
  {Graham}, G., {Dai}, Z.~R., {Chi}, M., {Bradley}, J., {Hagiya}, K.,
  {Gounelle}, M., {Bridges}, J., Feb. 2008. {Comparing Wild 2 particles to
  chondrites and IDPs}. Meteoritics and Planetary Science 43, 261--272.

\end{thebibliography}

\section*{Appendix}

	\begin{table*}
		\caption{Spectral feature fitting information part 1}             
		\centering                          
		\begin{tabular}  { l l l l l l l l l }       

\# & Name & Type & 16 \mic band & 24 \mic band & 9 \mic band & 15 \mic band & Py/(Ol+Py) & Ref. \\
\hline
\hline
1 & Charsonville & H 6 & 16.64+/-0.031 & 24.35+/-0.001 & 9.31+/-0.002 & 15.54+/-0.0 & 0.32+/-0.01 & - \\ 
2 & Hessle & H 5 & 16.72+/-0.004 & 24.39+/-0.001 & 9.39+/-0.001 & 15.54+/-0.001 & 0.41+/-0.01 & - \\ 
3 & Bremervorde & H 3 & 16.77+/-0.004 & 24.56+/-0.016 & 9.38+/-0.001 & 15.54+/-0.001 & 0.47+/-0.01 & - \\ 
4 & Pultusk & H 5 & 16.72+/-0.001 & 24.38+/-0.001 & 9.36+/-0.001 & 15.56+/-0.001 & 0.41+/-0.01 & - \\ 
5 & Menow & H 4 & 16.67+/-0.004 & 24.32+/-0.002 & 9.35+/-0.001 & 15.54+/-0.0 & 0.33+/-0.0 & - \\ 
6 & Stalldalen & H 5 & 16.69+/-0.001 & 24.31+/-0.001 & 9.33+/-0.001 & 15.55+/-0.001 & 0.36+/-0.01 & - \\ 
7 & Bjelaja & H 6 & 16.72+/-0.002 & 24.36+/-0.001 & 9.38+/-0.001 & 15.56+/-0.001 & 0.37+/-0.01 & - \\ 
8 & Ochansk & H 4 & 16.72+/-0.003 & 24.37+/-0.001 & 9.36+/-0.001 & 15.54+/-0.002 & 0.35+/-0.02 & - \\ 
\hline 
9 & Bjurbole & L 4 & 16.81+/-0.002 & 24.51+/-0.009 & 9.39+/-0.002 & 15.54+/-0.001 & 0.23+/-0.01 & - \\ 
\hline 
10 & Soka-Banja & LL 4 & 16.83+/-0.0 & 24.44+/-0.003 & 9.41+/-0.002 & 15.54+/-0.0 & 0.18+/-0.0 & - \\ 
11 & Ensisheim & LL 6 & 16.84+/-0.001 & 24.64+/-0.0 & 9.31+/-0.002 & 15.61+/-0.001 & 0.19+/-0.0 & - \\ 
\hline 
12 & Shalka & Dio  &  - &  - & 9.42+/-0.002 & 15.57+/-0.0 & 1.0+/-0.0 & - \\ 
13 & Juvinas & Euc  &  - &  - & 9.51+/-0.002 & 15.86+/-0.003 & 1.0+/-0.0 & - \\ 
14 & Bialystok & How  & 16.72+/-0.005 & 24.49+/-0.001 & 9.27+/-0.007 & 15.55+/-0.0 & 0.3+/-0.0 & - \\ 
15 & Luotolax & How  &  - &  - & 9.43+/-0.001 & 15.88+/-0.02 & 1.0+/-0.0 & - \\ 
16 & Stannern & Euc  &  - & 26.03+/-0.12 & 9.44+/-0.012 & 15.9+/-0.001 & 1.0+/-0.0 & - \\ 
\hline 
17 & Seymehan & Pal  & 16.46+/-0.001 & 24.05+/-0.001 &  - &  - & 0.0+/-0.0 & - \\ 
\hline 
18 & Cenniceros & H 3.8 & 16.82+/-0.001 & 24.37+/-0.001 &  - & 15.5+/-0.001 & 0.33+/-0.01 & 1 \\ 
\hline 
19 & Barratta & L 3.8 & 16.76+/-0.027 & 24.37+/-0.0 & 9.38+/-0.001 & 15.5+/-0.001 & 0.36+/-0.01 & 1 \\ 
\hline 
20 & Parnallee & LL 3.6 & 16.8+/-0.001 & 24.52+/-0.0 & 9.37+/-0.0 & 15.53+/-0.001 & 0.32+/-0.0 & 1 \\ 
\hline 
21 & Indarch & E 4 &  - &  - & 9.29+/-0.004 & 15.44+/-0.001 & 1.0+/-0.0 & 1 \\ 
\hline 
22 & Ave. Eucrite & Euc &  - &  - & 9.51+/-0.063 & 15.99+/-0.022 & 1.0+/-0.0 & 2 \\ 
23 & Ave. Diognite & Dio &  - &  - & 9.33+/-0.004 & 15.54+/-0.001 & 1.0+/-0.0 & 2 \\ 
24 & Ave. Howardite & How &  - &  - & 9.43+/-0.006 & 15.98+/-0.005 & 1.0+/-0.0 & 2 \\

\hline 

		\label{fittinginfo1}
		\end{tabular}
		Reference numbers indicate: "-": this work, 1: \cite{morlok14a}, 2: \cite{morlok12}, 3: \cite{chihara02}, 4:\cite{koike03}, 5: \cite{beck14}, 6: \cite{servoin73}, 7: \cite{jager98}, 8: \cite{salisbury91}
	\end{table*}

	\begin{table*}
		\caption{Spectral feature fitting information part 2}             
		\centering                          
		\begin{tabular}  { l l l l l l l l l}       

\# & Name & Type & 16 \mic band & 24 \mic band & 9 \mic band & 15 \mic band & Py/(Ol+Py) & Ref. \\
\hline 
\hline
31 & CEn100 & 0 &  - &  - & 9.28+/-0.011 & 15.41+/-0.005 & - & 3 \\ 
32 & CEn90 & 10 &  - &  - & 9.32+/-0.015 & 15.48+/-0.002 & - & 3 \\ 
33 & CEn80 & 20 &  - &  - & 9.32+/-0.004 & 15.5+/-0.002 & - & 3 \\ 
34 & CEn70 & 30 &  - &  - & 9.34+/-0.008 & 15.52+/-0.005 & - & 3 \\ 
35 & CEn60 & 40 &  - &  - & 9.37+/-0.01 & 15.58+/-0.001 & - & 3 \\ 
36 & CEn50 & 50 &  - &  - & 9.42+/-0.005 & 15.62+/-0.001 & - & 3 \\ 
37 & CEn0 & 100 &  - &  - & 9.5+/-0.001 & 15.91+/-0.005 & - & 3 \\ 
\hline 
38 & Fo100 & 0 & 16.34+/-0.001 & 23.74+/-0.011 &  - &  - & - & 4 \\ 
39 & Fo90 & 10 & 16.4+/-0.001 & 23.92+/-0.003 &  - &  - & - & 4 \\ 
40 & Fo84 & 16 & 16.61+/-0.013 & 24.35+/-0.001 &  - &  - & - & 4 \\ 
41 & Fo57 & 43 & 16.78+/-0.003 & 24.91+/-0.009 &  - &  - & - & 4 \\ 
42 & Fo40 & 60 & 17.12+/-0.002 & 25.77+/-0.011 &  - &  - & - & 4 \\ 
43 & Fo22 & 78 & 17.45+/-0.034 &  - &  - &  - & - & 4 \\ 
44 & Fo16 & 84 & 17.52+/-0.002 &  - &  - &  - & - & 4 \\ 
45 & Fo0 & 100 & 17.68+/-0.003 &  - &  - &  - & - & 4 \\ 
\hline 
46 & Fo GRF & 0.1 & 16.2+/-0.001 & 23.29+/-0.0 &  - &  - & - & 6 \\ 
47 & Fo GRF & 1.0 & 16.24+/-0.001 & 23.34+/-0.004 &  - &  - & - & 6 \\ 
48 & Fo GRF & 2.0 & 16.31+/-0.003 & 23.44+/-0.002 &  - &  - & - & 6 \\ 
49 & Fo GRF & 3.0 & 16.36+/-0.006 & 23.5+/-0.004 &  - &  - & - & 6 \\ 
50 & Fo GRF & 4.0 & 16.33+/-0.008 & 23.41+/-0.011 &  - &  - & - & 6 \\ 
51 & Fo GRF & 5.0 & 16.33+/-0.006 & 23.47+/-0.005 &  - &  - & - & 6 \\ 
\hline 
52 & En GRF & 0.1 &  - &  - & 9.17+/-0.011 & 15.41+/-0.005 & - & 7 \\ 
53 & En GRF & 1.0 &  - &  - & 9.21+/-0.001 & 15.42+/-0.001 & - & 7 \\ 
54 & En GRF & 2.0 &  - &  - & 9.26+/-0.029 & 15.42+/-0.007 & - & 7 \\ 
55 & En GRF & 3.0 &  - &  - & 9.19+/-0.0 & 15.42+/-0.009 & - & 7 \\ 
\hline 
		\label{fittinginfo2}
		\end{tabular}
	\end{table*}

	\begin{table*}
		\caption{Spectral feature fitting information part 3}             
		\centering                          
		\begin{tabular}  { l l l l l l l l l}       

\# & Name & Type & 16 \mic band & 24 \mic band & 9 \mic band & 15 \mic band & Py/(Ol+Py) & Ref. \\
\hline 
\hline 
56 & Albite &  - &  - &  - &  - &  - & - & 8 \\ 
57 & Anorthite &  - &  - &  - &  - &  - & - & 8 \\ 
\hline 
58 & ORGEUIL & CI 1 &  - &  - &  - &  - & - & 5 \\ 
59 & ALH83100 & CM 1 &  - &  - &  - &  - & - & 5 \\ 
60 & MET01070 & CM 1 &  - &  - &  - &  - & - & 5 \\ 
61 & GRO95577 & CR 1 &  - &  - &  - &  - & - & 5 \\ 
\hline 
62 & MURCHISON & CM 2 &  - &  - &  - &  - & - & 5 \\ 
63 & QUE99355 & CM 2 &  - &  - &  - &  - & - & 5 \\ 
64 & DOM03183 & CM 2 & 16.56+/-0.002 &  - &  - &  - & 0.0+/-0.0 & 5 \\ 
65 & WIS91600 & CM 2 & 16.35+/-0.001 & 23.76+/-0.002 &  - &  - & 0.0+/-0.0 & 5 \\ 
66 & LAP02342 & CR 2 & 16.37+/-0.002 & 23.64+/-0.002 & 9.34+/-0.001 & 15.44+/-0.002 & 0.58+/-0.03 & 5 \\ 
67 & MIL07700 & CM 2 & 16.49+/-0.001 & 23.85+/-0.001 &  - &  - & 0.0+/-0.0 & 5 \\ 
68 & MCY05230 & CM 2 & 16.37+/-0.001 &  - &  - & 15.38+/-0.001 & 0.69+/-0.01 & 5 \\ 
69 & EET83355 & C2 2 &  - & 23.75+/-0.006 & 9.33+/-0.001 &  - & - & 5 \\ 
70 & QUE97990 & CM 2 & 16.36+/-0.002 &  - &  - & 15.37+/-0.051 & 0.21+/-0.14 & 5 \\ 
71 & PCA91082 & CR 2 & 16.37+/-0.0 & 23.69+/-0.001 & 9.31+/-0.001 & 15.41+/-0.002 & 0.13+/-0.01 & 5 \\ 
72 & GRA06100 & CR 2 &  - & 23.68+/-0.011 & 9.31+/-0.001 &  - & - & 5 \\ 
73 & LON94101 & CM 2 & 16.35+/-0.001 &  - &  - &  - & 0.0+/-0.0 & 5 \\ 
74 & DOM03813 & CM 2 &  - &  - &  - &  - & - & 5 \\ 
75 & EET96029 & CM 2 &  - &  - &  - &  - & - & 5 \\ 
76 & DOM08003 & CM 2 &  - &  - &  - &  - & - & 5 \\ 
77 & PCA02010 & CM 2 & 16.83+/-0.001 &  - & 9.31+/-0.0 &  - & 0.0+/-0.0 & 5 \\ 
78 & ESSEBI & CM 2 & 16.35+/-0.0 &  - &  - &  - & 0.0+/-0.0 & 5 \\ 
79 & TAGISHLAKE & C2 2 & 16.3+/-0.0 &  - &  - &  - & 0.0+/-0.0 & 5 \\ 
80 & BORISKINO & CM 2 & 16.41+/-0.001 &  - &  - &  - & 0.0+/-0.0 & 5 \\ 
81 & PCA91008 & CM 2 &  - & 23.66+/-0.003 & 9.29+/-0.001 &  - & - & 5 \\ 
82 & GRO03116 & CR 2 &  - & 23.61+/-0.008 & 9.33+/-0.001 & 15.43+/-0.001 & 1.0+/-0.0 & 5 \\ 
83 & ALH84044 & CM 2 &  - &  - &  - &  - & - & 5 \\ 
84 & LEW85311 & CM 2 &  - &  - &  - &  - & - & 5 \\ 
85 & NOGOYA & CM 2 &  - &  - &  - &  - & - & 5 \\ 
86 & RBT04133 & CR 2 & 16.71+/-0.001 &  - & 9.33+/-0.001 & 15.37+/-0.001 & 0.08+/-0.0 & 5 \\ 
87 & LEW87022 & CM 2 &  - &  - &  - &  - & - & 5 \\ 
88 & NIGER & CM 2 &  - &  - &  - &  - & - & 5 \\ 
89 & BELLS & CM 2 & 16.36+/-0.001 &  - &  - &  - & 0.0+/-0.0 & 5 \\ 
\hline 
90 & ALLENDE & CV 3 & 16.78+/-0.002 &  - & 9.27+/-0.022 &  - & 0.0+/-0.0 & 5 \\ 
91 & KABA & CV 3 & 16.44+/-0.0 & 23.73+/-0.002 & 9.3+/-0.0 &  - & 0.0+/-0.0 & 5 \\ 
92 & GROSNAJA & CV 3 & 16.68+/-0.002 &  - & 9.31+/-0.002 &  - & 0.0+/-0.0 & 5 \\ 
93 & MOKOIA & CV 3 & 16.83+/-0.001 &  - & 9.29+/-0.002 &  - & 0.0+/-0.0 & 5 \\ 
94 & VIGARANO & CV 3 & 16.71+/-0.011 &  - & 9.31+/-0.001 & 15.36+/-0.002 & 0.04+/-0.0 & 5 \\ 
\hline 
95 & allende PO4 & chon. & 16.39+/-0.0 & 23.84+/-0.0 &  - &  - & 0.0+/-0.0 & 1 \\ 
96 & allende PO1 & chon. & 16.47+/-0.001 & 23.92+/-0.001 & 9.16+/-0.0 &  - & 0.0+/-0.0 & 1 \\ 
97 & allende POP3 & chon. & 16.39+/-0.001 & 23.85+/-0.0 & 9.23+/-0.0 &  - & 0.0+/-0.0 & 1 \\ 
98 & allende PO2 & chon. & 16.37+/-0.0 & 23.83+/-0.0 &  - &  - & 0.0+/-0.0 & 1 \\ 
\hline

		\label{fittinginfo3}
		\end{tabular}
	\end{table*}

	\begin{table*}
		\caption{Spectral feature continua points. Part 1 }             
		\label{table:1}      
		\centering                          
		\begin{tabular}  { l l l l l }       

\#  & 16 \mic band & 24 \mic band & 9 \mic band & 15 \mic band  \\
\hline 
\hline
1 & 15.9, 15.99, 17.2, 17.35 & 21.5, 22.32, 26.61, 27.34 & 8.65, 8.97, 9.67, 9.87 & 14.99, 15.11, 15.94, 16.02\\ 
2 & 15.9, 16.06, 17.3, 17.51 & 21.44, 22.55, 26.37, 28.21 & 8.55, 9.01, 9.65, 9.93 & 14.95, 15.06, 15.96, 16.11\\ 
3 & 15.93, 16.16, 17.26, 17.67 & 21.57, 22.68, 26.29, 26.93 & 8.46, 8.95, 9.66, 9.83 & 14.94, 15.09, 16.02, 16.22\\ 
4 & 15.94, 16.14, 17.3, 17.6 & 21.4, 22.44, 26.46, 27.62 & 8.65, 8.94, 9.69, 9.9 & 14.97, 15.12, 15.99, 16.21\\ 
5 & 15.9, 16.05, 17.28, 17.57 & 21.11, 22.27, 26.56, 27.42 & 8.65, 9.0, 9.66, 9.95 & 14.93, 15.09, 15.89, 16.08\\ 
6 & 15.93, 16.06, 17.32, 17.72 & 21.23, 22.4, 26.19, 27.5 & 8.56, 8.93, 9.73, 9.94 & 14.98, 15.1, 15.96, 16.11\\ 
7 & 15.98, 16.11, 17.28, 17.56 & 21.23, 22.33, 26.33, 27.86 & 8.68, 8.98, 9.65, 9.93 & 15.0, 15.19, 16.0, 16.19\\ 
8 & 15.96, 16.1, 17.32, 17.67 & 21.16, 22.58, 26.33, 27.42 & 8.59, 8.93, 9.69, 9.95 & 15.0, 15.16, 15.92, 16.11\\ 
9 & 15.93, 16.1, 17.41, 17.79 & 21.16, 22.45, 26.21, 27.49 & 8.84, 8.97, 9.63, 9.85 & 14.99, 15.1, 15.95, 16.12\\ 
10 & 15.89, 16.04, 17.53, 17.76 & 21.5, 22.38, 26.72, 28.26 & 8.78, 9.02, 9.69, 9.95 & 15.07, 15.15, 15.94, 16.13\\ 
11 & 15.93, 16.19, 17.43, 17.8 & 21.42, 22.26, 26.83, 27.88 & 8.65, 8.97, 9.65, 9.81 & 15.1, 15.26, 16.03, 16.21\\ 
12 &  -  &  -  & 8.44, 8.64, 10.07, 10.21 & 15.01, 15.09, 16.3, 16.64\\ 
13 &  -  &  -  & 8.9, 9.08, 9.98, 10.17 & 15.19, 15.34, 16.41, 16.72\\ 
14 & 15.9, 16.04, 17.29, 17.5 & 21.29, 22.2, 26.6, 27.87 & 8.6, 8.94, 9.67, 9.78 & 14.96, 15.09, 15.93, 16.13\\ 
15 &  -  &  -  & 8.56, 8.85, 9.97, 10.14 & 15.16, 15.26, 16.72, 17.01\\ 
16 &  -  & 22.63, 23.09, 28.08, 28.23 & 8.66, 9.16, 10.03, 10.41 & 15.27, 15.46, 16.32, 16.67\\ 
17 & 14.86, 15.44, 17.22, 17.77 & 20.9, 21.98, 26.1, 27.3 &  -  &  - \\ 
18 & 15.95, 16.06, 17.38, 17.9 & 21.48, 22.51, 26.4, 27.16 &  -  & 14.85, 15.1, 15.95, 16.15\\ 
19 & 15.91, 16.11, 17.35, 17.84 & 21.73, 22.46, 26.55, 27.17 & 8.87, 9.06, 9.71, 9.87 & 14.86, 15.15, 15.96, 16.14\\ 
20 & 15.92, 16.12, 17.37, 17.72 & 21.39, 22.46, 26.5, 27.74 & 8.79, 8.99, 9.73, 9.84 & 14.9, 15.06, 16.03, 16.17\\ 
21 &  -  &  -  & 8.5, 8.78, 9.62, 10.0 & 14.71, 14.97, 15.97, 16.39\\ 
22 &  -  &  -  & 8.86, 9.02, 10.05, 10.23 & 15.13, 15.39, 16.5, 16.79\\ 
23 &  -  &  -  & 8.76, 9.01, 9.79, 10.05 & 14.8, 15.14, 16.22, 16.77\\ 
24 &  -  &  -  & 8.74, 8.93, 10.05, 10.24 & 15.11, 15.4, 16.44, 16.74\\ 
31 &  -  &  -  & 8.52, 8.87, 9.65, 9.76 & 14.89, 15.11, 15.79, 16.21\\ 
32 &  -  &  -  & 8.29, 8.84, 9.64, 9.78 & 14.92, 15.1, 15.96, 16.22\\ 
33 &  -  &  -  & 8.26, 8.84, 9.73, 9.87 & 14.86, 14.94, 15.86, 16.39\\ 
34 &  -  &  -  & 8.18, 8.74, 9.82, 10.14 & 14.94, 15.24, 16.2, 16.44\\ 
35 &  -  &  -  & 8.31, 8.69, 9.95, 10.13 & 15.04, 15.14, 16.03, 16.47\\ 
36 &  -  &  -  & 8.35, 8.68, 9.97, 10.12 & 15.14, 15.24, 16.2, 16.65\\ 
37 &  -  &  -  & 9.15, 9.25, 9.64, 9.71 & 15.45, 15.53, 16.52, 17.3\\ 
38 & 14.1, 15.27, 16.87, 17.45 & 20.98, 22.01, 25.6, 26.77 &  -  &  - \\ 
39 & 14.82, 15.43, 17.09, 17.64 & 21.5, 22.15, 25.64, 27.79 &  -  &  - \\ 
40 & 13.97, 15.38, 17.34, 18.0 & 20.92, 21.97, 26.23, 27.62 &  -  &  - \\ 
41 & 14.84, 15.8, 17.5, 18.0 & 21.3, 22.25, 26.58, 27.38 &  -  &  - \\ 
42 & 14.97, 15.88, 17.83, 18.51 & 23.77, 24.44, 27.44, 28.57 &  -  &  - \\ 
43 & 15.54, 16.46, 18.12, 18.63 &  -  &  -  &  - \\ 
44 & 15.61, 16.49, 18.35, 18.76 &  -  &  -  &  - \\ 
45 & 15.49, 16.6, 18.6, 19.12 &  -  &  -  &  - \\ 
46 & 13.9, 15.25, 16.91, 17.43 & 21.2, 21.92, 24.22, 24.6 &  -  &  - \\ 
47 & 14.12, 15.12, 16.99, 17.43 & 21.5, 21.82, 24.34, 24.85 &  -  &  - \\ 
48 & 14.39, 15.24, 17.0, 17.54 & 21.26, 21.77, 24.59, 25.15 &  -  &  - \\ 
49 & 13.86, 15.27, 17.02, 17.59 & 21.34, 21.89, 24.59, 24.96 &  -  &  - \\ 
50 & 14.35, 15.29, 17.1, 17.5 & 21.26, 22.2, 24.49, 24.82 &  -  &  - \\ 
51 & 13.52, 14.88, 17.19, 17.59 & 21.09, 21.97, 24.45, 24.84 &  -  &  - \\ 
52 &  -  &  -  & 8.33, 8.78, 9.51, 10.05 & 14.6, 15.01, 15.74, 16.1\\ 
53 &  -  &  -  & 8.75, 8.89, 9.48, 9.73 & 14.6, 15.0, 15.74, 16.06\\ 
54 &  -  &  -  & 8.74, 8.93, 9.5, 9.76 & 14.6, 15.02, 15.8, 16.44\\ 
55 &  -  &  -  & 8.79, 8.9, 9.48, 9.65 & 14.74, 15.02, 15.93, 16.28\\ 
56 &  -  &  -  &  -  &  - \\ 
57 &  -  &  -  &  -  &  - \\ 
58 &  -  &  -  &  -  &  - \\ 
59 &  -  &  -  &  -  &  - \\ 
60 &  -  &  -  &  -  &  - \\ 

		\label{tab_cont_points}
		\end{tabular}
	\end{table*}

	\begin{table*}
		\caption{Spectral feature continua points. Part 2}             
		\label{table:1}      
		\centering                          
		\begin{tabular}  { l l l l l }       

\#  & 16 \mic band & 24 \mic band & 9 \mic band & 15 \mic band  \\
\hline 
\hline
61 &  -  &  -  &  -  &  - \\ 
62 &  -  &  -  &  -  &  - \\ 
63 &  -  &  -  &  -  &  - \\ 
64 & 15.53, 15.92, 17.15, 17.68 &  -  &  -  &  - \\ 
65 & 15.53, 15.78, 17.05, 17.46 & 22.65, 23.15, 24.45, 24.99 &  -  &  - \\ 
66 & 15.83, 16.02, 16.74, 17.11 & 21.87, 22.37, 24.56, 24.9 & 8.77, 8.9, 9.64, 9.8 & 14.85, 15.04, 15.84, 16.07\\ 
67 & 14.96, 15.58, 17.65, 18.29 & 22.48, 22.95, 24.63, 24.95 &  -  &  - \\ 
68 & 15.8, 15.99, 17.02, 17.3 &  -  &  -  & 14.66, 14.82, 15.89, 16.11\\ 
69 &  -  & 21.92, 22.63, 25.27, 26.12 & 8.84, 9.02, 9.65, 9.73 &  - \\ 
70 & 15.52, 15.84, 16.74, 17.25 &  -  &  -  & 14.94, 15.03, 15.7, 15.91\\ 
71 & 15.43, 15.68, 17.03, 17.34 & 21.29, 22.24, 24.58, 24.88 & 8.79, 8.96, 9.56, 9.69 & 14.9, 15.07, 15.85, 16.08\\ 
72 &  -  & 21.76, 22.4, 24.78, 26.04 & 8.43, 8.89, 9.67, 9.73 &  - \\ 
73 & 15.63, 15.97, 16.8, 17.09 &  -  &  -  &  - \\ 
74 &  -  &  -  &  -  &  - \\ 
75 &  -  &  -  &  -  &  - \\ 
76 &  -  &  -  &  -  &  - \\ 
77 & 15.39, 15.9, 17.73, 18.22 &  -  & 8.62, 8.94, 9.58, 9.71 &  - \\ 
78 & 15.54, 15.91, 16.69, 16.91 &  -  &  -  &  - \\ 
79 & 15.47, 15.82, 16.66, 16.87 &  -  &  -  &  - \\ 
80 & 15.37, 15.92, 16.99, 17.45 &  -  &  -  &  - \\ 
81 &  -  & 21.93, 22.53, 24.42, 24.94 & 8.78, 8.91, 9.57, 9.67 &  - \\ 
82 &  -  & 21.95, 22.62, 24.67, 24.99 & 8.73, 8.93, 9.58, 9.77 & 14.94, 15.11, 16.01, 16.12\\ 
83 &  -  &  -  &  -  &  - \\ 
84 &  -  &  -  &  -  &  - \\ 
85 &  -  &  -  &  -  &  - \\ 
86 & 15.62, 15.95, 17.92, 18.31 &  -  & 8.73, 8.88, 9.68, 9.81 & 14.78, 14.99, 15.86, 16.16\\ 
87 &  -  &  -  &  -  &  - \\ 
88 &  -  &  -  &  -  &  - \\ 
89 & 15.41, 15.84, 17.01, 17.63 &  -  &  -  &  - \\ 
90 & 15.43, 15.93, 17.8, 18.18 &  -  & 8.68, 8.98, 9.69, 9.78 &  - \\ 
91 & 15.01, 15.44, 17.44, 17.87 & 22.25, 22.61, 24.59, 25.08 & 8.79, 8.95, 9.56, 9.68 &  - \\ 
92 & 15.24, 15.5, 17.84, 18.07 &  -  & 8.65, 9.03, 9.63, 9.79 &  - \\ 
93 & 15.33, 15.71, 17.92, 18.32 &  -  & 8.72, 8.96, 9.61, 9.68 &  - \\ 
94 & 15.76, 15.94, 17.69, 17.98 &  -  & 8.7, 8.89, 9.58, 9.71 & 14.87, 15.01, 15.73, 15.94\\ 
95 & 14.8, 15.29, 17.22, 17.63 & 21.32, 21.84, 26.3, 26.7 &  -  &  - \\ 
96 & 15.04, 15.41, 17.28, 17.76 & 21.37, 22.01, 26.01, 26.62 & 8.6, 8.88, 9.55, 9.81 &  - \\ 
97 & 15.14, 15.3, 17.05, 17.42 & 21.51, 21.94, 26.03, 26.74 & 8.58, 8.87, 9.57, 9.72 &  - \\ 
98 & 14.55, 15.14, 17.25, 17.7 & 21.35, 21.91, 26.02, 26.73 &  -  &  - \\ 

		\label{tab_cont_points2}
		\end{tabular}
	\end{table*}

\end{document}